\documentclass[letterpaper,11pt]{article}
\usepackage{graphicx,amssymb,amstext,amsmath,amsfonts}
\usepackage{fullpage}
\usepackage{cite}
\usepackage{bm}
\usepackage{multicol}

\makeatletter
\def\@cite#1#2{\textsuperscript{{#1\if@tempswa , #2\fi}}}
\makeatother

\makeatletter
\def\@biblabel#1{#1.}
\makeatother

\newenvironment{methods}{%
    \section*{Methods}%
    \setlength{\parskip}{0pt}%
    }{}

\newcommand{\onlinecite}[1]{\hspace{-1 ex} \nocite{#1}\citenum{#1}}

\newcommand{\spacing}[1]{\renewcommand{\baselinestretch}{#1}\large\normalsize}
\spacing{2} 

\renewcommand{\figurename}{{\bf{Figure}}}

\def\@maketitle{%
  \newpage\spacing{1}\setlength{\parskip}{12pt}%
  {\Large\bfseries\noindent\sloppy \textsf{\@title} \par}%
    {\noindent\@author}%
}

\newenvironment{affiliations}{%
    \setcounter{enumi}{1}%
    \setlength{\parindent}{0in}%
    \slshape\sloppy%
    \begin{list}{\upshape$^{\arabic{enumi}}$}{%
        \usecounter{enumi}%
        \setlength{\leftmargin}{0in}%
        \setlength{\topsep}{0in}%
        \setlength{\labelsep}{0in}%
        \setlength{\labelwidth}{0in}%
        \setlength{\listparindent}{0in}%
        \setlength{\itemsep}{0ex}%
        \setlength{\parsep}{0in}%
        }
    }{\end{list}\par\vspace{12pt}}

\renewenvironment{abstract}{%
    \setlength{\parindent}{0in}%
    \setlength{\parskip}{0in}%
    \sffamily\bfseries%
    }{\par\vspace{-6pt}}

\begin{document}

\noindent\parbox{\textwidth}{\center{
\textsf{\textbf{%
\LARGE{Two-stage magnetic-field-tuned superconductor-insulator transition in underdoped La$_{\bm{2-x}}$Sr$_{\bm{x}}$CuO$_{\bm{4}}$}
\vspace{7mm}
}}}}
\noindent\parbox{\textwidth}{\center{
\textsf{\textbf{%
Xiaoyan Shi$^{\bm 1 \dag}$, Ping V. Lin$^{\bm 1}$, T. Sasagawa$^{\bm 2}$, V. Dobrosavljevi\'{c}$^{\bm 1}$ \&  Dragana Popovi\'{c}$^{\bm {1*}}$
\vspace{7mm}
}}}}
\noindent\parbox{\textwidth}{\center
\today
\vspace{7mm}
}
\begin{affiliations}
 \item National High Magnetic Field Laboratory and Department of Physics, Florida State University, Tallahassee, Florida 32310, USA
 \item Materials and Structures Laboratory, Tokyo Institute of Technology, Kanagawa 226-8503, Japan
\end{affiliations}
\begin{itemize}
  \item $^{\dag}$ Present address: Sandia National Laboratories, Albuquerque, New Mexico 87185, USA  \vspace*{-12pt}
  \item * E-mail: dragana@magnet.fsu.edu
\end{itemize}

\newpage

\begin{abstract}
In the underdoped pseudogap regime of cuprate superconductors, the normal state is commonly probed by applying a magnetic field (${\bm H}$).  However, the nature of the
${\bm H}$-induced resistive state has been the subject of a long-term debate, and clear evidence for a zero-temperature (${\bm{T=0}}$) ${\bm H}$-tuned
superconductor-insulator transition (SIT) has proved elusive.  Here we report magnetoresistance measurements in underdoped
La$_{\bm{2-x}}$Sr$_{\bm x}$CuO$_{\bm 4}$, providing striking evidence for quantum critical behavior of the resistivity -- the signature of a ${\bm H}$-driven SIT.
The transition is not direct: it is accompanied by the emergence of an intermediate state, 
which is a superconductor only at ${\bm{T=0}}$.  Our finding of a two-stage ${\bm H}$-driven SIT goes beyond the conventional scenario in which a single quantum critical point separates the superconductor and the insulator in the presence of a perpendicular ${\bm H}$.
Similar two-stage ${\bm H}$-driven SIT, in which both disorder and quantum phase fluctuations play an important role, may also be expected in other copper-oxide high-temperature superconductors.
\end{abstract}

The SIT is an example of a quantum phase transition (QPT): a continuous phase transition that occurs at  $T=0$, controlled by some parameter of the Hamiltonian of the system, such as doping or the external magnetic field\cite{Sachdev-book}.  A QPT can affect the behavior of the system up to surprisingly high temperatures.  In fact, many unusual properties of various strongly correlated materials have been attributed to the proximity of quantum critical points (QCPs).  An experimental signature of a QPT at nonzero $T$ is the observation of scaling behavior with relevant parameters in describing the data.    Although the SIT has been studied extensively\cite{Gantmakher-rev}, even in conventional superconductors many questions remain about the perpendicular-field-driven SIT in two-dimensional (2D) or quasi-2D systems\cite{CIQPT}.  In high-$T_c$ cuprates ($T_c$ -- transition temperature), which have a quasi-2D nature, early magnetoresistance (MR) experiments showed the suppression of superconductivity with high $H$, revealing the insulating behavior\cite{Ando-1995,Marta1996,GSB-1996} and hinting at an underlying $H$-field-tuned SIT\cite{Steiner-PRL2005}.  However, even though understanding the effects of $H$ is believed to be essential to understanding high-$T_c$
cuprate superconductivity and continues to be a subject of intensive research, the evidence for the $H$-field-driven SIT and the associated QPT scaling in cuprates remains scant and inconclusive.

In the conventional picture of type-II superconductors, $H$
penetrates the sample in the form of a solid lattice of interacting vortex lines in the entire mixed  state $H_{c1}(T)<H<H_{c2}(T)$, where $H_{c1}$ is the Meissner field and $H_{c2}$ is the upper critical field.  This picture, however, neglects fluctuations which, in high-$T_c$ superconductors, are especially important.   Indeed, the delicate interplay of thermal fluctuations, quantum fluctuations and disorder leads to a complex $H-T$ phase diagram of the vortex matter\cite{Blatter,Rosenstein-rev,Doussal-rev}.  Thermal fluctuations, for example, cause melting of the vortex solid into a vortex liquid for fields below what is now a crossover line $H_{c2}(T)$.  Quantum fluctuations could result in a vortex liquid persisting down to $T=0$.  At very low $T$, the disorder becomes important and modifies the vortex phase diagram such that there are two distinct vortex solid phases\cite{Rosenstein-rev,Doussal-rev}: a Bragg glass with $T_c(H)> 0$ at lower fields and a vortex glass (VG) with $T_c=0$ at higher fields.  The SIT would then correspond to a transition from a VG to an insulator at even higher $H$.  However, the interplay of this vortex line physics and quantum criticality, the key question in the $H$-field-tuned SIT, has remained largely unexplored.

In this study, we find strong evidence for the $H$-field-driven SIT in underdoped La$_{2-x}$Sr$_{x}$CuO$_4$ (LSCO).  The results are consistent with the existence of three phases at $T=0$, although the behavior of the in-plane resistivity $\rho$ suggests the presence of the direct SIT over a surprisingly wide range of $T$ and $H$.  At the lowest $T$, however, $\rho(T,H)$ reveal an intermediate phase with $T_c=0$ and the true SIT at higher $H$.  We focus on samples with a relatively low $T_c(H=0)$ to ensure that experimentally available $H$ are high enough to fully suppress superconductivity.  Unlike most other studies, ours includes samples grown using different techniques (Methods), in order to separate out any effects that may depend on the sample preparation conditions from the more general behavior.  
In addition to providing evidence for the SIT, the MR data are used to calculate the contribution of superconducting fluctuations (SCFs) to conductivity, allowing a construction of the $H-T$ phase diagram.

\noindent \textbf{In-plane resistivity of La$_{\bm{2-x}}$Sr$_{\bm x}$CuO$_{\bm 4}$}\\
One sample was a film with the nominal composition La$_{1.93}$Sr$_{0.07}$CuO$_4$ (ref. \onlinecite{XShi-NatMat}) and a measured $T_c(H=0)=(3.8\pm 0.1)$~K.  The high-quality single crystal\cite{SasagawaLSCO} had $T_{c}(H=0)=(5.2\pm 0.1)$~K and the nominal composition La$_{1.94}$Sr$_{0.06}$CuO$_{4+y}$, with $y$ not precisely known (see Methods for  more details).  Unless otherwise specified, $T_c$ is defined throughout as the temperature at which the resistance (\textit{i.e.} $\rho$) becomes zero for a given $H$.  (The method to determine $T_c(H)$ is illustrated in Supplementary Fig.~S1.)

Figure~\ref{fig:rhoTH}a shows the $\rho(T)$ curves for different $H$ that were extracted from the MR measurements at
\begin{figure}
\centerline{\normalsize{\textbf a}\includegraphics[width=11cm]{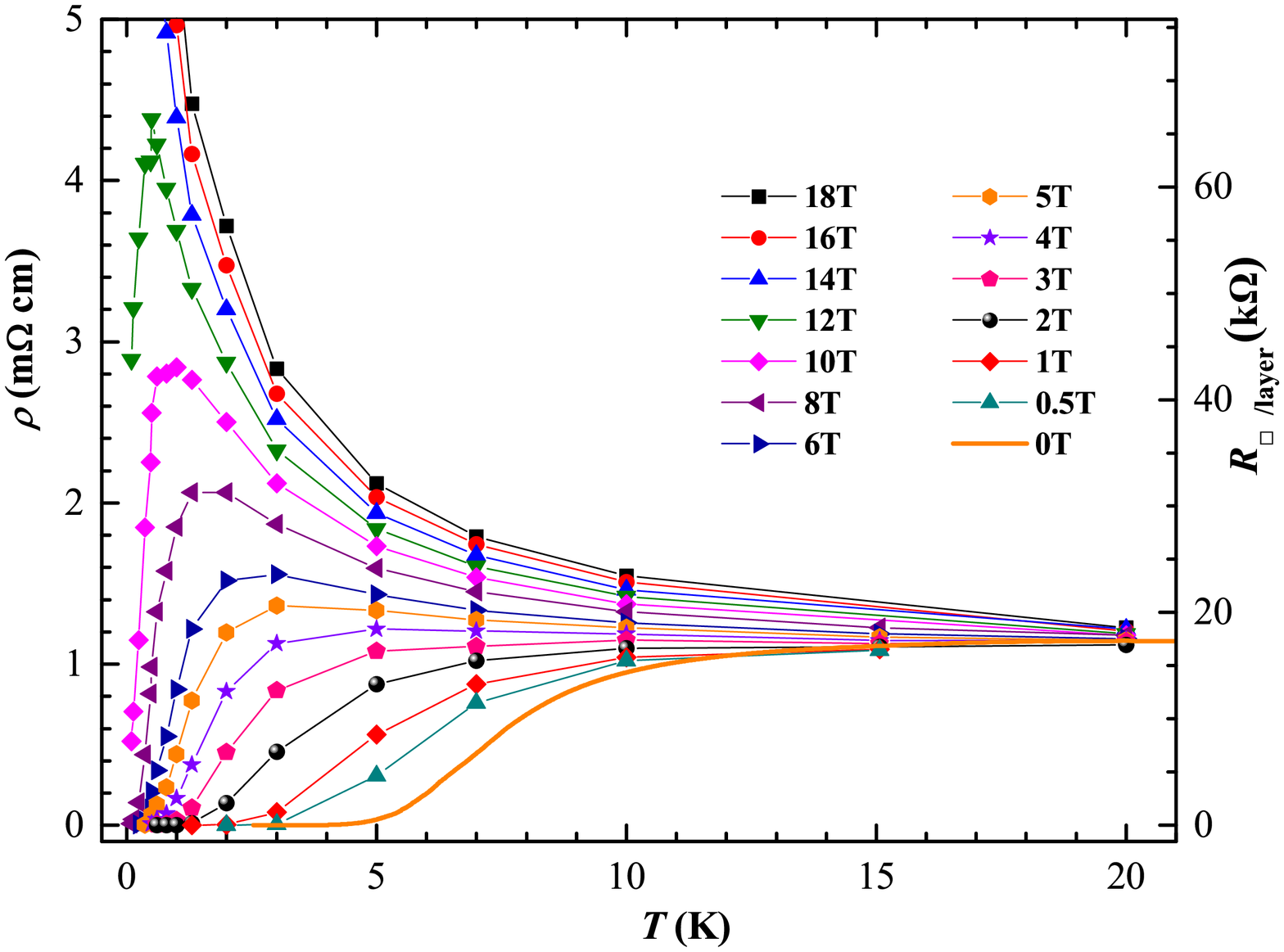}}\vspace*{0.5cm}
\centerline{\normalsize{\textbf b}\includegraphics[width=8cm]{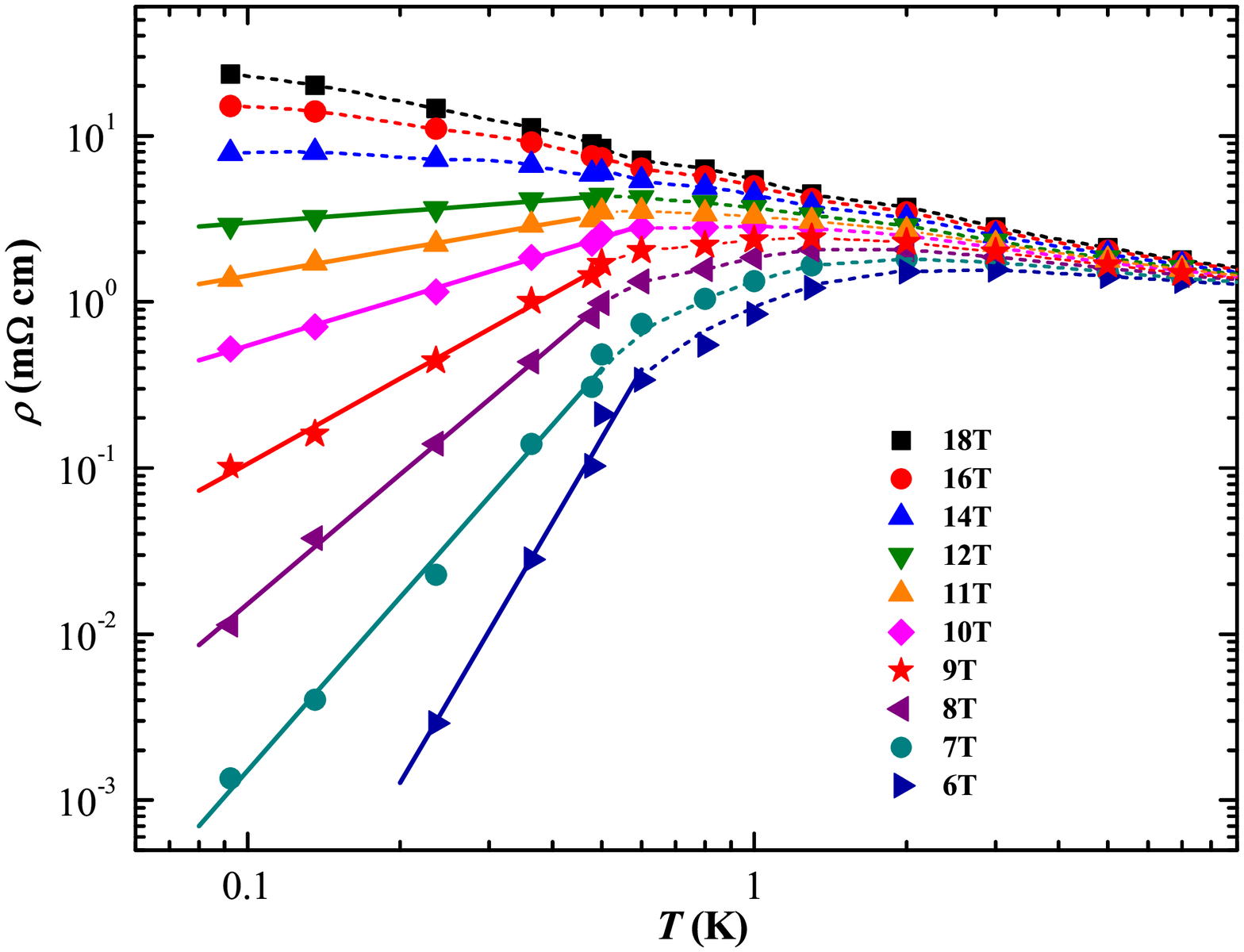}
\normalsize{\textbf c}\includegraphics[width=8.5cm]{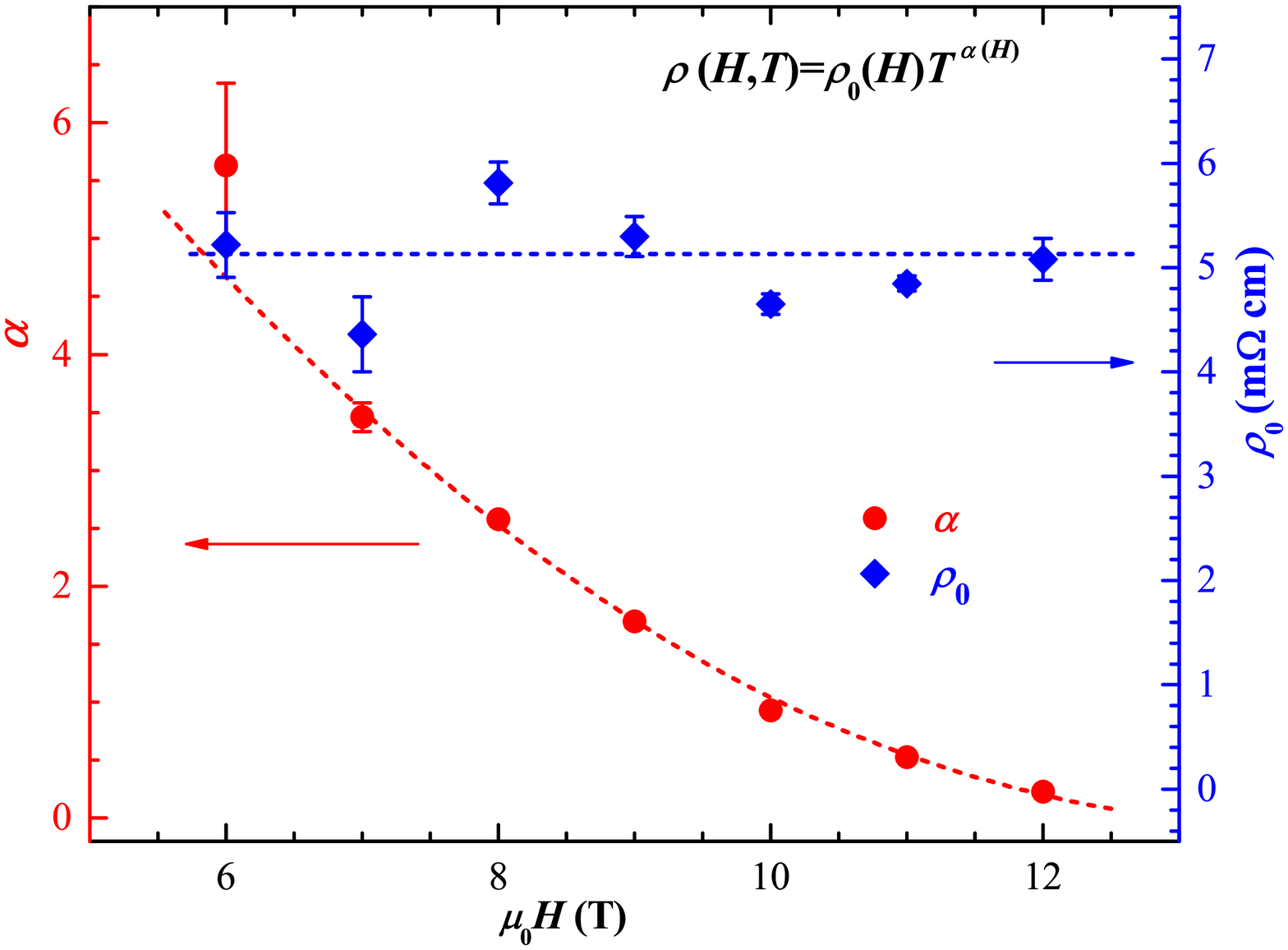}}
\caption{\textbf{Temperature dependence of the in-plane resistivity ${\bm{\rho}}$ in different magnetic fields ${\bm{H\parallel c}}$ for ${\bm{x=0.07}}$ LSCO film.}
 \textbf{a,} The $\rho(T,H)$ data exhibit a change of the sign of $d\rho/dT$ as a function of $H$ at high $T\gtrsim5$~K ($R_{\square/layer}$ is resistance per square per CuO$_2$ layer; see Methods).  Except for $H=0$, solid lines guide the eye.  \textbf{b,} Some of the $\rho(T,H)$ in \textbf{a} are shown on a log-log scale to focus on the low-$T$, intermediate-$H$ regime.  Short-dashed lines guide the eye.  Solid lines represent power-law fits $\rho(T,H)=\rho_{0}(H)T^{\alpha(H)}$.  \textbf{c,} Fitting parameters $\rho_{0}(H)$ and $\alpha(H)$.  Error bars indicate 1 standard deviation (s.d.) in the fits for $\rho_{0}(H)$ and $\alpha(H)$.  Short-dashed lines guide the eye.}
\label{fig:rhoTH}
\end{figure}
fixed $T$ for the $x=0.07$ sample (see Supplementary Fig.~S2 for similar data on the $x=0.06$ crystal).  Small $H$
clearly lead to a decrease of $T_c(H)$, such that $T_c\rightarrow 0$ for $H\approx 4$~T (Supplementary Information and Fig.~S3).  The survival of the superconducting phase with $T_c> 0$ up to fields much higher than the Meissner field $H_{c1}\sim 100$~Oe is understood to be a consequence of the pinning of vortices by disorder\cite{Blatter,Rosenstein-rev,Doussal-rev}.  As $H$ increases further, a pronounced maximum appears in $\rho(T)$.  The temperature at the maximum, $T_{max}$, shifts to lower $T$ with increasing $H$, similar to early observations\cite{Marta1996}.  At the highest $H$, $\rho(T)$ curves exhibit insulating behavior.

The intermediate $H$ regime in which each $\rho(T)$ curve exhibits a maximum is specially intriguing.  Here the system shows a tendency towards insulating behavior already at high $T\gtrsim5$~K, but then the sign of $d\rho/dT$ changes, suggesting that another mechanism sets in at lower $T$ and drives $\rho(T\rightarrow 0)$ towards zero.  The low-$T$, intermediate-$H$ regime is more evident in Fig.~\ref{fig:rhoTH}b, where the same data are presented on a log-log scale.  For $T<T_{max}$, the data are described best with the phenomenological power-law fits $\rho(T,H)=\rho_{0}(H)T^{\alpha(H)}$, which indicate that $\rho(T= 0)=0$, \textit{i.e.} that the system is a superconductor only at $T=0$.  The exponent $\alpha$ depends on $H$ (Fig.~\ref{fig:rhoTH}c), and goes to zero at $H\sim 13.5$~T.  This implies the existence of a $T$-independent $\rho$ at that field.  The non-monotonic behavior of $\rho(T)$ in the intermediate $H$ regime suggests that high-$T$ ($T>T_{max}$) and low-$T$ ($T<T_{max}$) regions should be examined separately and more closely.

\begin{figure}
\centerline{\normalsize{\textbf a}\includegraphics[width=11cm]{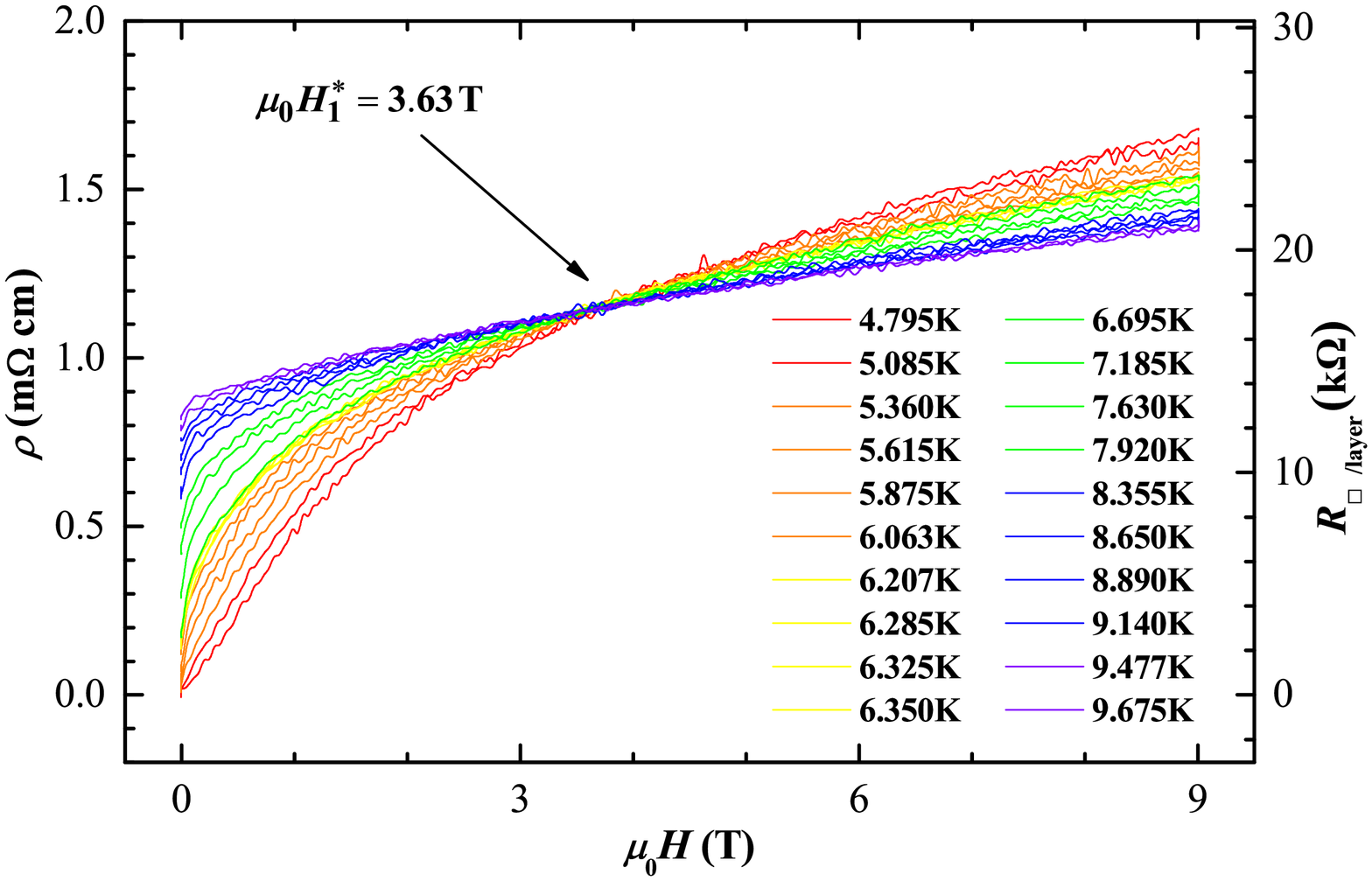}}\vspace*{0.5cm}
\begin{minipage}[t]{0.5\textwidth} \normalsize{\textbf b}\includegraphics[width=8.6cm]{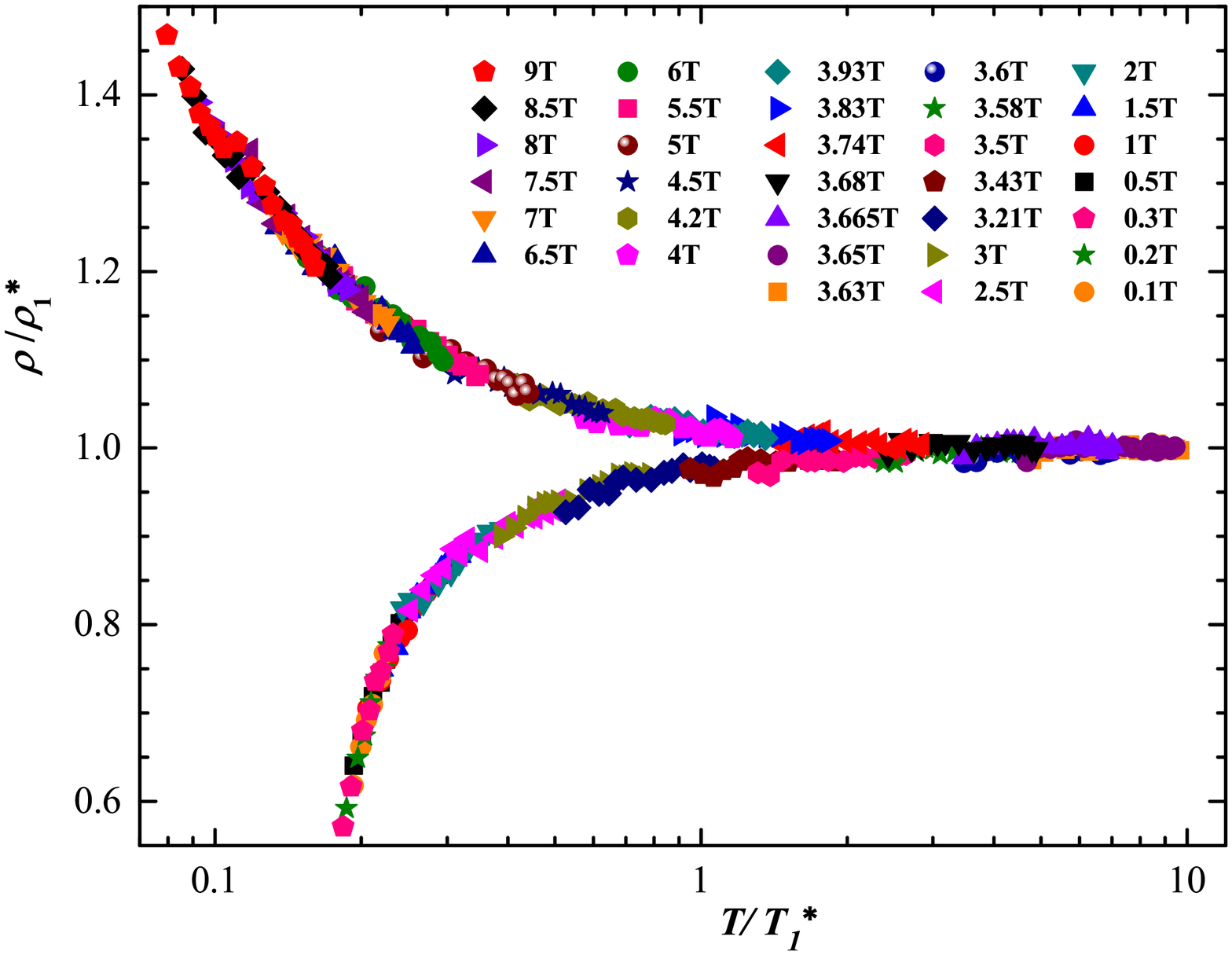}
\end{minipage}
\begin{minipage}[t]{0.5\textwidth} \hspace*{0.5cm}\normalsize{\textbf c}\includegraphics[width=8cm]{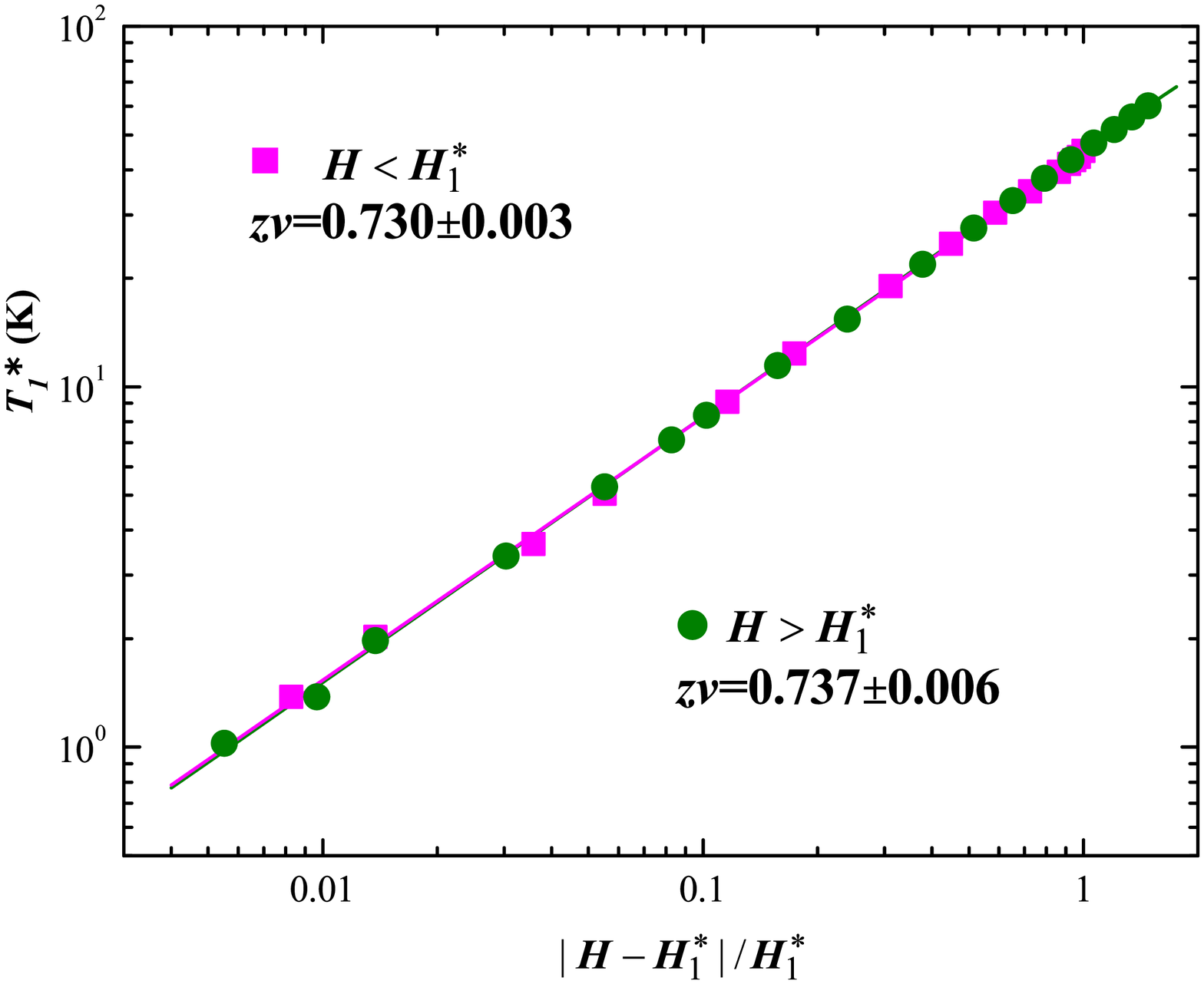}
\end{minipage}
\caption{\textbf{High-temperature ($T\gtrsim5$~K) behavior of the resistivity ${\bm{\rho}}$ in different magnetic fields ${\bm{H\parallel c}}$ for ${\bm{x=0.07}}$ LSCO film.}
 \textbf{a,} Isothermal $\rho(H)$ curves in the high-$T$ region show the existence of a $T$-independent crossing point at $\mu_{0}H_{1}^{\ast}=3.63$~T and $\rho_{1}^{\ast}=1.15$~m$\Omega\cdot$cm (or $R_{\square/layer}\approx 17.4$~k$\Omega$).  \textbf{b,} Scaling of the data in \textbf{a} with respect to a single variable $T/T_{1}^{\ast}$.  The scaling region is shown in more detail in Fig.~\ref{fig:phase}a.  \textbf{c,} The scaling parameter $T_{1}^{\ast}$ as a function of $\mid H-H_{1}^{\ast}\mid/H_{1}^{\ast}$ on both sides of $H_{1}^{\ast}$.  The lines are fits with slopes $z\nu\approx 0.73$, as shown.}
\label{fig:highT}
\end{figure}

\noindent \textbf{Scaling analysis of the high-temperature behavior}\\
Figure~\ref{fig:highT}a shows a more detailed set of MR measurements on the $x=0.07$ film focused on the behavior at 5~K~$\lesssim T< 10$~K.
The MR curves clearly exhibit a well-defined, $T$-independent crossing point at $\mu_{0}H_{1}^{\ast}=3.63$~T.  In other words, this is the field where $d\rho/dT$ changes sign from positive or metallic at low $H$, to negative or insulating at $H>H_{1}^{\ast}$ (see also Fig.~\ref{fig:rhoTH}a).  We find that, near $H_{1}^{\ast}$, $\rho(T)$ for different $H$ can be collapsed onto a single function by rescaling the temperature, as shown in Fig.~\ref{fig:highT}b.  Therefore, $\rho(T,H)=\rho_{1}^{\ast}f_{1}(T/T_{1}^{\ast})$, where the scaling parameter $T_{1}^{\ast}$ is found to be the same function of $\delta=(H-H_{1}^{\ast})/H_{1}^{\ast}$ on both sides of $H_{1}^{\ast}$ .  In particular, $T_{1}^{\ast}\propto |\delta|^{z\nu}$ with $z\nu\approx 0.73$ (Fig.~\ref{fig:highT}c) over a remarkably wide, more than two orders of magnitude range of $|\delta|$.  Such a single-parameter scaling of the resistance is precisely what is expected\cite{MPAF-SIT} near a $T=0$ SIT in 2D, where $z$ and $\nu$ are the dynamical and correlation length exponents, respectively\cite{Gantmakher-rev}.  Similar scaling on the $x=0.06$ crystal sample (Supplementary Fig.~S4) yields a comparable $z\nu=0.59\pm 0.08$.  It is interesting that, although the critical fields $H_{1}^{\ast}$ in the two samples differ by almost a factor of two ($\mu_{0}H_{1}^{\ast}=3.63$~T for $x=0.07$ film and $\mu_{0}H_{1}^{\ast}=6.68$~T for $x=0.06$ crystal), the critical resistivities $\rho_{1}^{\ast}$ are almost the same ($R_{\square/layer}\approx (17.4-18.0)$~k$\Omega$; see also Supplementary Information).  The exponent product $z\nu\sim 0.7$ has been observed for perpendicular $H$-field-tuned transitions also in conventional 2D superconductors (\textit{e.g.} in a-Bi\cite{Markovic} and a-NbSi\cite{Aubin}) and, more recently, in 2D superconducting LaTiO$_3$/SrTiO$_3$ interfaces\cite{Biscaras}.  The value $z\nu\sim 0.7$ is believed to be in the universality class of the 2D SIT in the clean limit, as described by the $(2+1)$D~$XY$ model and assuming that $z=1$ due to the long-range Coulomb interaction between charges\cite{MPAF-SIT,MPAF-GG}.

\begin{figure}
\centerline{\normalsize{\textbf a}\includegraphics[width=10cm]{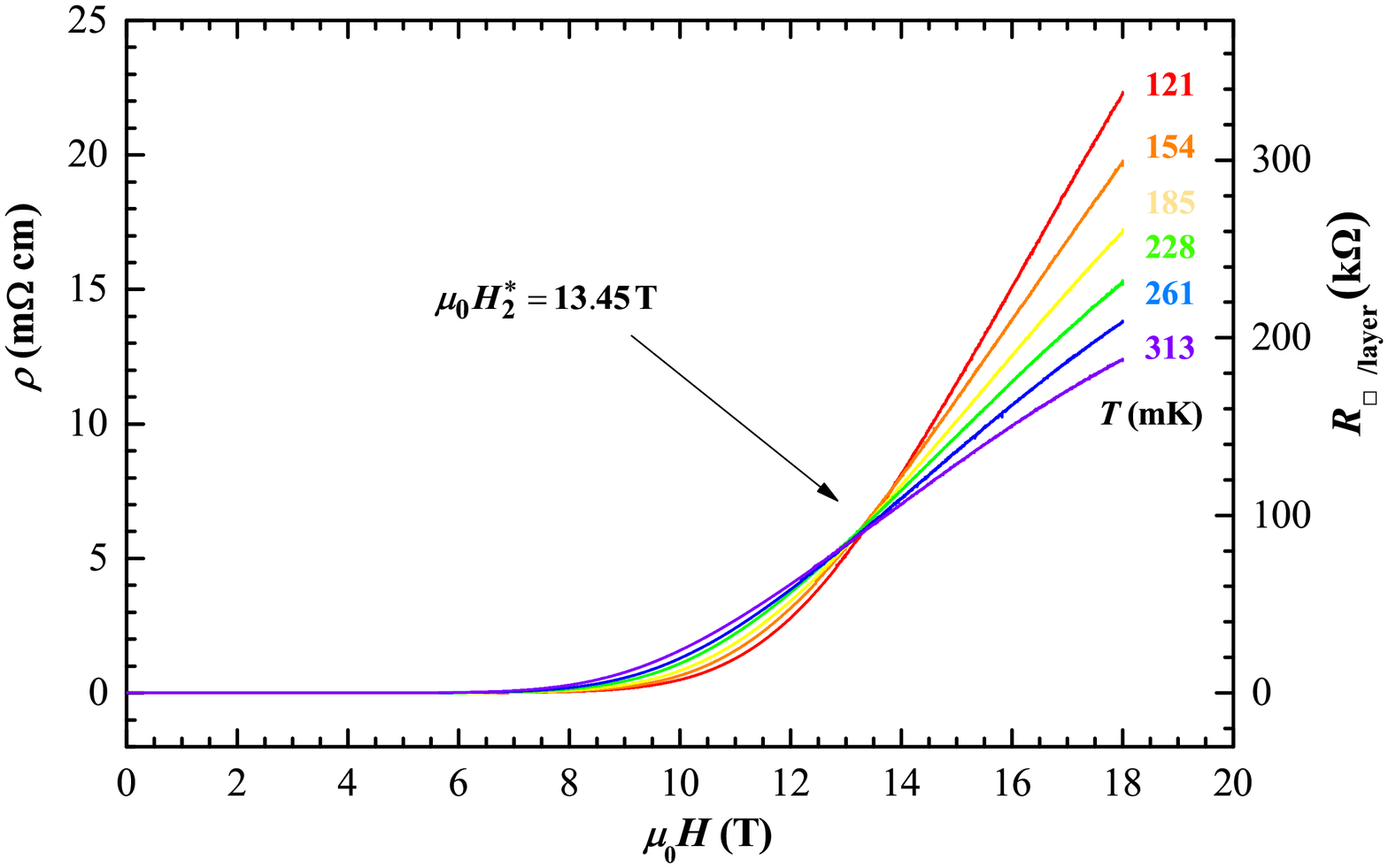}}\vspace*{0.5cm}
\begin{minipage}[t]{0.5\textwidth} \normalsize{\textbf b}\includegraphics[width=8.5cm]{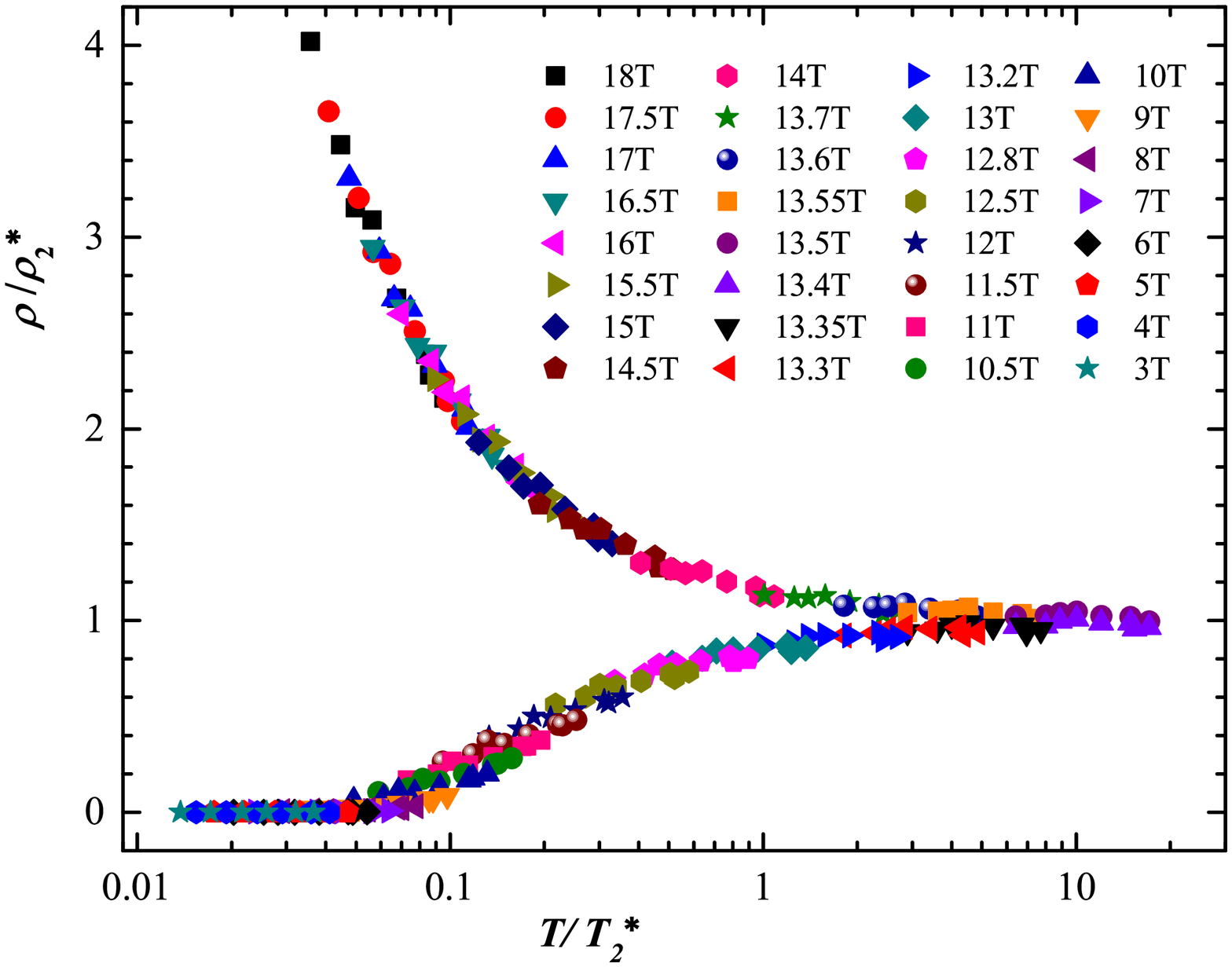}
\end{minipage}
\begin{minipage}[t]{0.5\textwidth} \hspace*{0.5cm}\normalsize{\textbf c}\includegraphics[width=8cm]{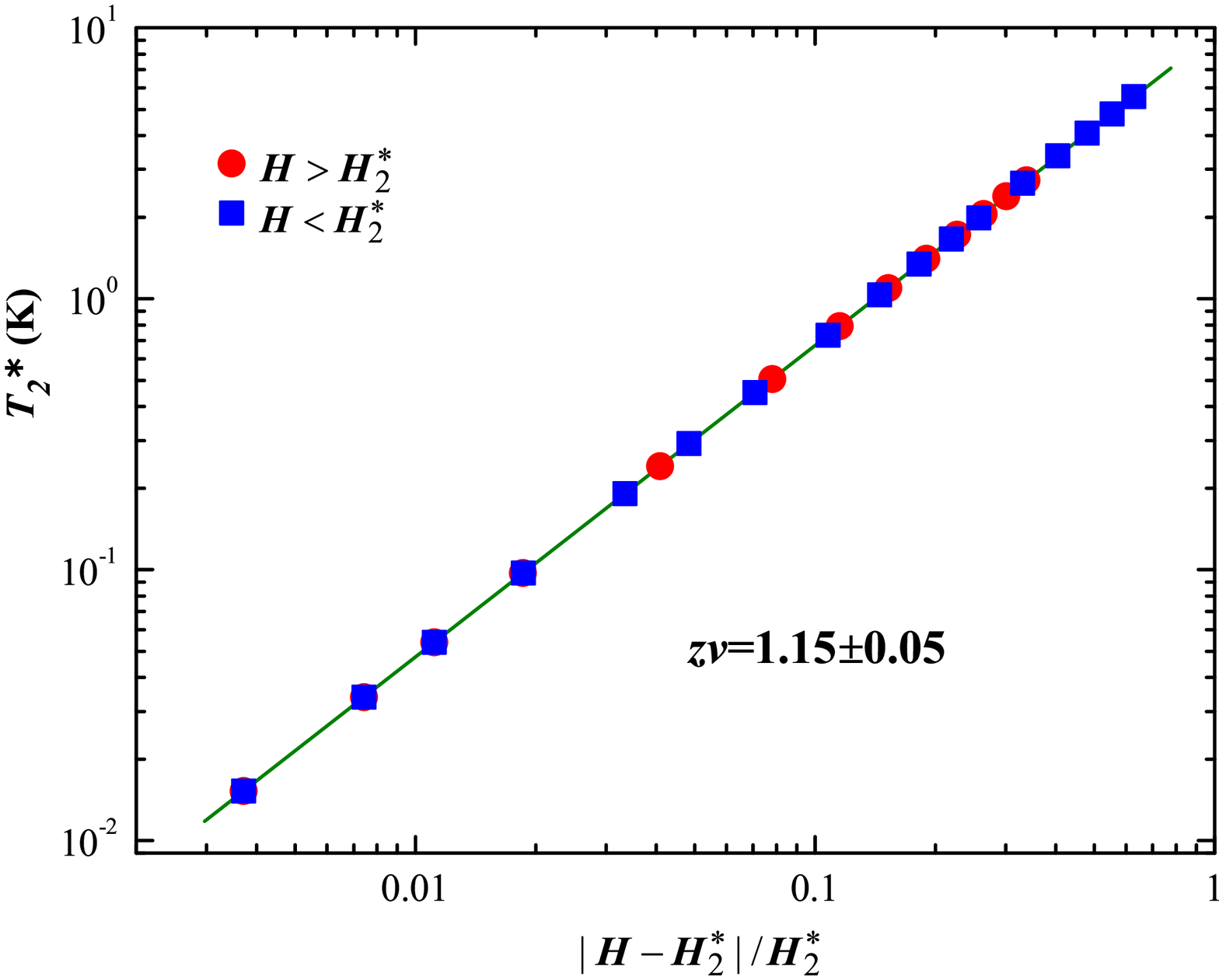}
\end{minipage}
  \caption{\textbf{Low-temperature ($T\lesssim 0.3$~K) behavior of the resistivity ${\bm{\rho}}$ in different magnetic fields ${\bm{H\parallel c}}$ for ${\bm{x=0.07}}$ LSCO film.}  \textbf{a,}  Isothermal $\rho(H)$ curves in the low-$T$ region show the existence of a $T$-independent crossing point at $\mu_{0}H_{2}^{\ast}=13.45$~T and $\rho_{2}^{\ast}=6.404$~m$\Omega\cdot$cm (or $R_{\square/layer}\approx 97$~k$\Omega$).  \textbf{b,} Scaling of the data in \textbf{a} with respect to a single variable $T/T_{2}^{\ast}$.  The scaling region, which is shown in more detail in Fig.~\ref{fig:phase}a,  includes the data at the lowest $T\approx 0.09$~K.  \textbf{c,} The scaling parameter $T_{2}^{\ast}$ as a function of $\mid H-H_{2}^{\ast}\mid/H_{2}^{\ast}$ on both sides of $H_{2}^{\ast}$.  The line is a fit with slope $z\nu\approx 1.15$.}
  \label{fig:lowT}
\end{figure}

\noindent \textbf{Scaling analysis of the low-temperature behavior}\\
As shown above, the behavior of the system over a range of $T$ above $T_{max}$ appears to be controlled by the 
QCP corresponding to the transition from a superconductor to an insulator in the absence of disorder, and driven by quantum phase fluctuations.  As $T$ is lowered below $T_{max}$, however, this transition does not actually take place as some of the insulating curves assume the power-law dependence $\rho(T,H)=\rho_{0}(H)T^{\alpha(H)}$ (Fig.~\ref{fig:rhoTH}), leading to a superconducting state at $T=0$.  Figure~\ref{fig:lowT}a shows a set of MR measurements carried out at very low $T<T_{max}$, which exhibit a $T$-independent crossing point at $\mu_{0}H_{2}^{\ast}=13.45$~T, consistent with $\alpha(H=H_{2}^{\ast})\approx 0$ (see also Figs.~\ref{fig:rhoTH}b,c).  Near $H_{2}^{\ast}$, an excellent scaling of $\rho$ with temperature according to $\rho(T,H)=\rho_{2}^{\ast}f_{2}(T/T_{2}^{\ast})$ is obtained (Fig.~\ref{fig:lowT}b), where $T_{2}^{\ast}\propto |\delta|^{z\nu}$ on both sides of $H_{2}^{\ast}$ (Fig.~\ref{fig:lowT}c).  Here
$\delta=(H-H_{2}^{\ast})/H_{2}^{\ast}$ and $z\nu=1.15\pm 0.05$.  Assuming $z=1$, this type of single-parameter scaling with $\nu \geqslant 1$ corresponds to the $T=0$ SIT in a 2D disordered system\cite{MPAF-SIT}.

\noindent \textbf{Superconducting fluctuations}\\
The extent of SCFs can be determined from the transverse ($H\parallel c$) MR by mapping out fields $H_{c}'(T)$ above which the normal state is fully restored\cite{YBCO-SCF,YBCO-SCFlong,LSCO-SCF,XShi-NatMat}.  In the normal state at low fields, the MR increases as $H^2$ (ref.~\onlinecite{Harris}), so that the values of $H_{c}'$ can be found from the downward deviations from such quadratic dependence that arise from superconductivity when $H<H_{c}'$.  
The $H_{c}'(T)$ line determined using this method (see Supplementary Fig.~S5) is shown in Fig.~\ref{fig:phase}a for the $x=0.07$
\begin{figure}
\centerline{\normalsize{\textbf a} \includegraphics[width=10cm]{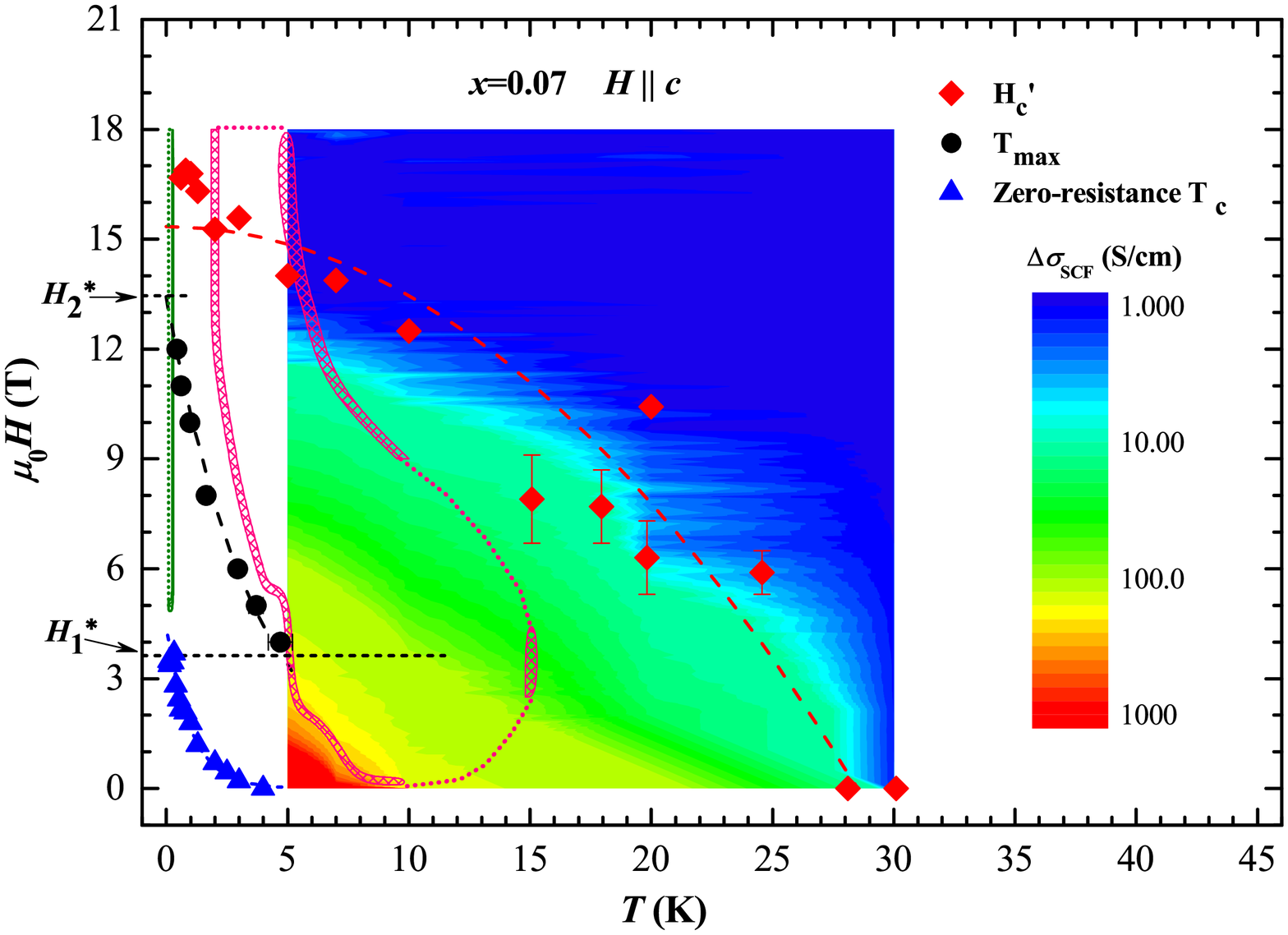}}
\centerline{\normalsize{\textbf b}\includegraphics[width=10cm]{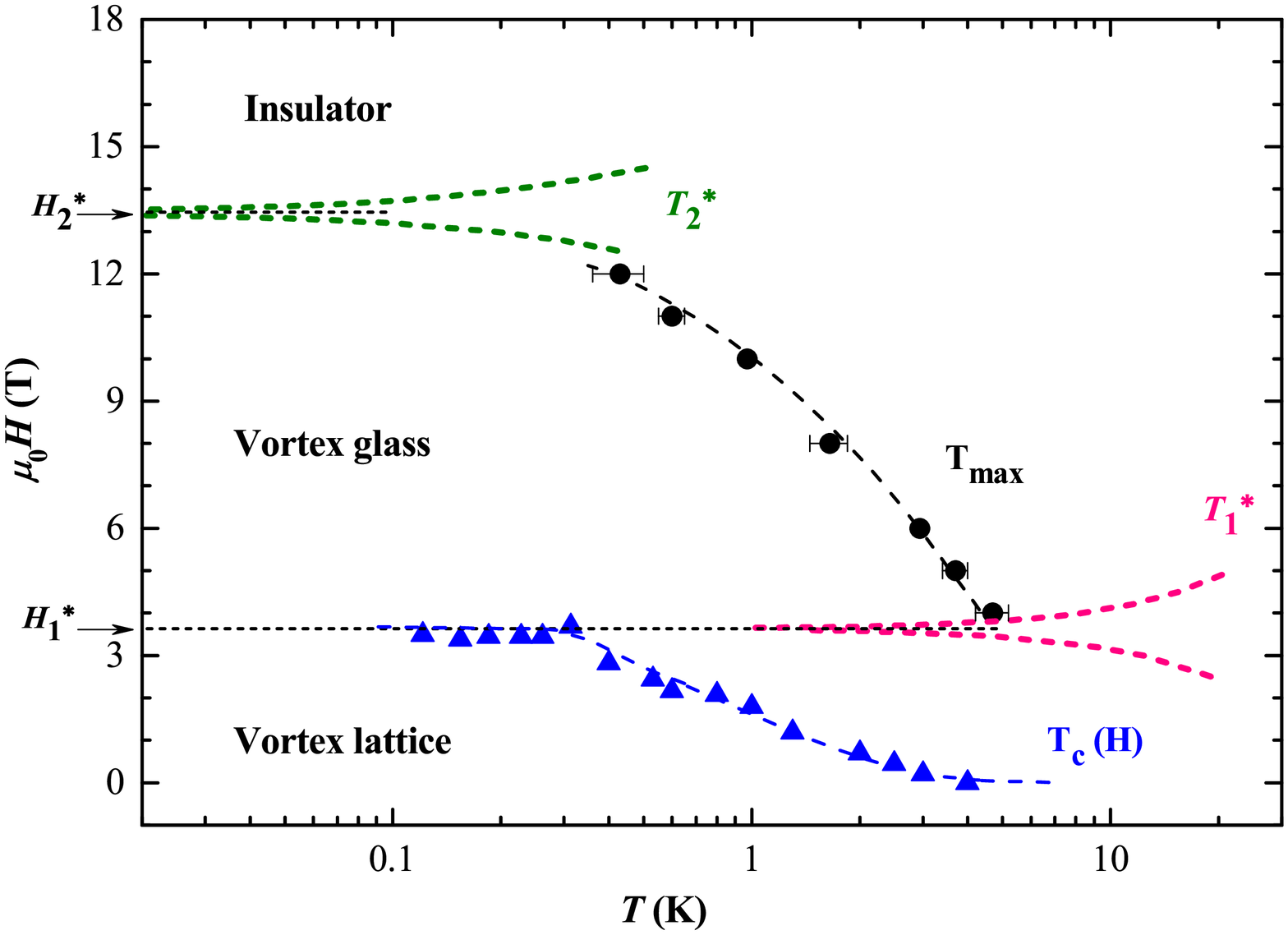}}
\caption{\textbf{Transport ${\bm{H-T}}$ phase diagram and scaling regions in underdoped LSCO.}  The data are shown for $x=0.07$ film.  \textbf{a},  The color map (on a log scale) shows the SCF contribution to conductivity $\Delta\sigma_{SCF}$ \textit{vs.} $T$ and $H\parallel c$.  The dashed red line is a fit with $\mu_{0}H_{c}^{'}$[T]$=15[1-(T$[K]$/29)^{2}]$.  The error bars indicate the uncertainty in $H_{c}^{'}$ that corresponds to 1 s.d. in the slopes of linear fits in Supplementary Fig. S5.  The horizontal dashed black lines mark the values of the $T=0$ critical fields $H_{1}^{\ast}$ and $H_{2}^{\ast}$ for scaling.  The pink lines show the high-$T$ scaling region: the hashed symbols mark areas beyond which scaling fails, and the dots indicate areas beyond which the data are not available.  The green lines show the low-$T$ scaling region.  \textbf{b},  Simplified phase diagram showing the crossover temperatures $T_{1}^{\ast}$ and $T_{2}^{\ast}$ corresponding to $H_{1}^{\ast}$ and $H_{2}^{\ast}$, respectively, and three phases at $T=0$.  The error bars for $T_{max}$ indicate 3 s.d. from fitting $\rho(T)$ in Fig.~\ref{fig:rhoTH}a.}
  \label{fig:phase}
\end{figure}
film\cite{XShi-NatMat} (see Supplementary Information and Fig.~S6 for the $x=0.06$ crystal sample).  In both cases, $H_{c}'(T)$ is well fitted by $H_{c}'(T)=H_{c}'(0)[1-(T/T_2)^2]$.  Figure~\ref{fig:phase}a also shows the SCF contribution to the conductivity\cite{YBCO-SCF,YBCO-SCFlong} $\Delta\sigma_{SCF}(T,H)=\rho^{-1}(T,H)-\rho_{n}^{-1}(T,H)$, where the normal-state resistivity $\rho_{n}(T,H)$ was obtained by extrapolating the region of $H^2$ magnetoresistance observed at high enough $H$ and $T$, as illustrated in Supplementary Fig.~S5.

\noindent \textbf{Phase diagram}\\
In addition to $H_{c}'(T)$ and $\Delta\sigma_{SCF}(T,H)$, the phase diagram in Fig.~\ref{fig:phase}a includes $T_c(H)$ and $T_{max}(H)$, as well as the critical fields $H_{1}^{\ast}$ and $H_{2}^{\ast}$.  For both samples, $T_c(H=H_0)\rightarrow 0$ (Supplementary Information and Fig.~S3) for fields $H_0\approx H_{1}^{\ast}$ ($\mu_{0}H_{1}^{\ast}=3.63$~T for $x=0.07$ film and $\mu_{0}H_{1}^{\ast}=6.68$~T for $x=0.06$ crystal).  This is consistent with the $T=0$ transition from a pinned vortex solid to another phase at higher fields.  For the $x=0.07$ film sample, $T_{max}(H=H_0)\rightarrow 0$ for $\mu_{0}H_0=(13.4\pm 0.7)$~T, \textit{i.e.} $T_{max}$ vanishes at the critical field $\mu_{0}H_{2}^{\ast}=13.45$~T within the measurement error.  In the $x=0.06$ single crystal, $T_{max}(H)$ extrapolates to zero at $\mu_{0}H_0=(14.0\pm 0.6)$~T, \textit{i.e.} at a field similar to that in the film sample, even though their $H_{c}'(T=0)$ are very different.  The vanishing of a characteristic energy scale, such as $T_{max}(H)$, in the $T=0$ limit is consistent with the existence of a quantum phase transition at $H_{2}^{\ast}$.

The scaling regions associated with the critical fields $H_{1}^{\ast}$ and $H_{2}^{\ast}$ are also shown in  Fig.~\ref{fig:phase}a.  It is striking that the ``hidden'' critical point at $H_{1}^{\ast}$, corresponding to the SIT in the clean limit, dominates a huge part of the phase diagram.  (See also Supplementary Fig.~S7 for scaling of $\rho(T,H)$ in the $x=0.07$ film over the entire scaling region shown in Fig.~\ref{fig:phase}a.)  For $H<H_{1}^{\ast}$, this scaling fails at low $T$ precisely where SCFs increase dramatically, leading to a large drop in $\rho$.
For $H>H_{1}^{\ast}$, the scaling works, of course, only for $T>T_{max}$.  The tendency towards insulating behavior that is observed already at low $H\gtrsim H_{1}^{\ast}$ and $T\gtrsim 5$~K (Fig.~\ref{fig:rhoTH}a) clearly belongs to the same family of insulating curves that span the fields up to $H\approx H_{c}'$ at low $T$.  It is apparent from Fig.~\ref{fig:highT} (also Supplementary Fig.~S7) that the insulating $\rho(T)$ does not follow the $\ln T$ dependence\cite{Ando-1995}.  The data are instead consistent (not shown) with the variable-range hopping (VRH), although the range of available fields is too small here to observe an orders-of-magnitude change in $\rho$ that is characteristic of VRH.  A larger change was observed in some early studies\cite{Marta1996,Marta1997} of the $H$-field induced localization in underdoped LSCO with a similar $T_c(H=0)$, and attributed to a 2D or 3D Mott VRH.

The scaling region corresponding to the critical point at $H_{2}^{\ast}$, on the other hand, is remarkably small (Fig.~\ref{fig:phase}a), but it dominates the behavior at the lowest $T$.  On the insulating side, \textit{i.e.} for $H>H_{2}^{\ast}$, there is evidence for the presence of SCFs, as the applied $H\lesssim H_{c}'$.  The lowest-$T$ insulating behavior (Fig.~\ref{fig:lowT}b) may also be fitted to a VRH form, but the change of $\rho$ is too small again to determine the VRH exponent with high certainty.

A simplified version of the same phase diagram is shown in Fig.~\ref{fig:phase}b, which includes crossover temperatures $T_{1}^{\ast}$ and $T_{2}^{\ast}$ corresponding to the two QCPs
$H_{1}^{\ast}$ and $H_{2}^{\ast}$, respectively.  The logarithmic $T$-scale also makes it more apparent that the vortex solid phase with $T_c(H)> 0$ extends precisely up to $H_{1}^{\ast}$, \textit{i.e.} that the QCP
$H_{1}^{\ast}$ is associated with the quantum melting of the pinned vortex solid.  The other phases shown in Fig.~\ref{fig:phase}b are discussed below.

\begin{figure}
\centerline{\includegraphics[width=15cm]{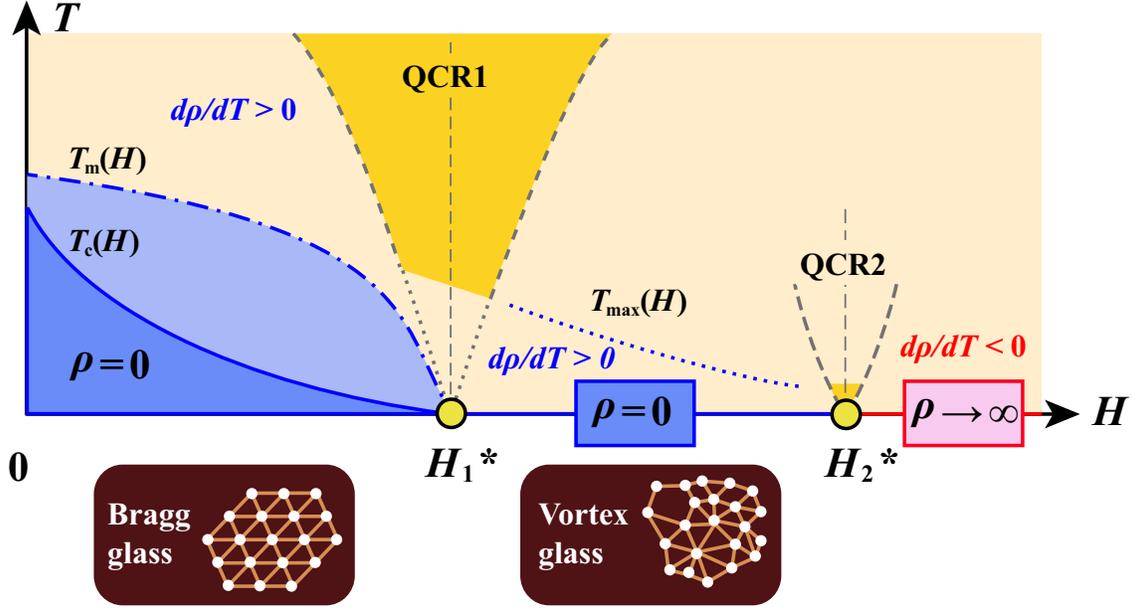}}
\caption{\textbf{Sketch of the interplay of vortex physics and quantum critical behavior in the ${\bm{H-T}}$ phase diagram.}   Two critical fields, $H_{1}^{\ast}$ and $H_{2}^{\ast}$, are reported, separating three distinct phases at $T=0$: i) a superconductor with $\rho=0$ (dark blue) for all $T<T_c(H)$ [$T_c(H)>0$] and $H<H_{1}^{\ast}$, ii) a superconductor with $\rho=0$ only at $T=0$ (\textit{i.e.} $T_c=0$) for $H_{1}^{\ast}<H<H_{2}^{\ast}$, and iii) an insulator (red), where $\rho(T=0)\rightarrow\infty$, for $H_{2}^{\ast}<H$.   The difference between the two superconducting ground states is in the ordering of the vortex matter: a pinned vortex solid (Bragg glass) for $H<H_{1}^{\ast}$ and a vortex glass for $H>H_{1}^{\ast}$, as shown schematically.   The quantum critical regions (QCRs) corresponding to $H_{1}^{\ast}$ and $H_{2}^{\ast}$ are also shown schematically (dashed lines).  The QC scaling associated with $H_{1}^{\ast}$ does not extend to the lowest $T$ (see dotted lines); apparently, this QPT is ``hidden'' at low $T$ by thermal fluctuations that cause the melting of the pinned vortex solid (Bragg glass) at $T_m(H)$ for $H<H_{1}^{\ast}$, and by the effects of disorder for $H>H_{1}^{\ast}$.  $T_{max}$ is the temperature at which $d\rho/dT$ changes sign.}
  \label{fig:sketch}
\end{figure}

\noindent \textbf{Discussion}\\
To account for our observation of three distinct phases as $T\rightarrow 0$ and two QPTs, we discuss our results in the context of other relevant work on the same material (see Fig.~\ref{fig:sketch} for a sketch of the phase diagram, which is based mainly on Figs.~\ref{fig:phase} and S6).  In particular, we note that two order parameters are needed for a complete 
description of a type-II superconductor in $H$:
one that describes superconductivity and another one that describes vortex matter (see, \textit{e.g.}, ref. \onlinecite{Zlatko}).

At low fields below the $T_c(H)$ line, $\rho(T)=0$, which is attributed to the pinning of the vortex solid\cite{Blatter,Rosenstein-rev,Doussal-rev}.  $T_c(H)$ is known\cite{S2,S3} to correspond to the so-called irreversibility temperature $T_{irr}(H)$ below which the magnetization becomes hysteretic, indicating a transition of the vortex system between a low-$T$, low-$H$ pinned regime and an unpinned one\cite{Doussal-rev}.  This low-$H$ vortex solid phase has been identified experimentally as a Bragg glass in LSCO (ref. \onlinecite{Bragg-LSCO}), as well as in other cuprates\cite{Rosenstein-rev,Doussal-rev} and some conventional superconductors (\textit{e.g.} $2H$-NbSe$_2$ (ref. \onlinecite{Andrei-PRL})).  The Bragg glass forms when the disorder is weak\cite{TG-PD-1,TG-PD-2}: it retains the topological order of the Abrikosov vortex lattice (see sketch of a Bragg glass in Fig.~\ref{fig:sketch}) but yields broadened diffraction peaks.  Since such a distorted Abrikosov lattice has many metastable states and barriers to motion, it is, strictly speaking, a glass.

At higher $H$,
where the density of vortices is larger\cite{TG-PD-1,TG-PD-2}, a topologically disordered, amorphous vortex glass 
is expected (see sketch of a VG in Fig.~\ref{fig:sketch}).  A transition from a Bragg glass to a VG with increasing $H$ has been observed in LSCO (ref. \onlinecite{Bragg-LSCO}) and other cuprates\cite{Rosenstein-rev,Doussal-rev},
consistent with our conclusion about two distinct superconducting phases at $T=0$: a superconductor with $T_c(H)> 0$ for $H<H_{1}^{\ast}$ and a superconductor with $T_c=0$ for $H_{1}^{\ast}<H<H_{2}^{\ast}$.  The power-law $\rho(T)$ observed for $T<T_{max}$ and $H>H_{1}^{\ast}$ is indeed reminiscent of the behavior expected in a vortex liquid above the glass transition\cite{MPAF-VG89,Fisher-Fisher-VG,MPAF-VG92} occurring at $T_g=0$,
consistent with theoretical considerations that found the VG phase to be unstable at non-zero temperature\cite{VGTg-1,VGTg-2} even in 3D.

In general, the theoretical description of the $T=0$ melting of the vortex lattice (Bragg glass in Fig.~\ref{fig:sketch}), associated with the QCP
at $H_{1}^{\ast}$ (see Fig.~\ref{fig:phase}b), remains an open problem.  However, it is known that, at finite $T$, melting of the vortex lattice by proliferation of dislocations corresponds to a phase transition in the 2D $XY$ model\cite{Rosenstein-rev,Doussal-rev}.  It is thus plausible that the (2+1)D $XY$ model in the clean limit could describe the $T=0$ melting of the pinned vortex solid, consistent with our findings.  The QC scaling
associated with this QPT does not extend all the way down to the lowest $T$.  For $H>H_{1}^{\ast}$, the data suggest that the effects of disorder become
important at low $T$,
causing the freezing of the vortex liquid into a VG phase with $T_c=0$.   $T_{max}(H)$ may represent a crossover energy scale associated with the glassy freezing of vortices, although there is some evidence in other cuprates that a similar line is a continuous (second-order) glass transition\cite{Rosenstein-rev,Doussal-rev}.
For $H<H_{1}^{\ast}$, the scaling no longer works at low $T$ where $\rho$ exhibits a large drop (Fig.~\ref{fig:phase}a), \textit{i.e.} for $T$ lower than a field-dependent temperature scale resembling the $T_m(H)$ line sketched in Fig.~\ref{fig:sketch}.  It is known that a sharp drop in $\rho$ is associated with a jump in the reversible magnetization\cite{S3}, which is indicative of the thermal melting of the low-field vortex solid phase\cite{S3,Rosenstein-rev,Doussal-rev}.  In general, the thermal melting takes place at $T_{m}(H)>T_{irr}(H)$. Therefore, the failure of scaling for $H<H_{1}^{\ast}$ on the low-$T$ side (Fig.~\ref{fig:phase}a) may be attributed to the melting of the pinned state.  Indeed, our observation of QC
scaling above the non-zero melting temperature is consistent with general expectations\cite{Sachdev-book}.

At even higher $H$, the transition from the superconducting VG phase to an insulator with localized Cooper pairs occurs at $H_{2}^{\ast}$ (where $H_{1}^{\ast}<H_{2}^{\ast}\lesssim H_{c}'$), consistent with boson-dominated models of the SIT (ref. \onlinecite{MPAF-SIT}) and earlier arguments\cite{Steiner-PRL2005}.  The QC
behavior associated with this QPT is observed down to the lowest measured $T$.

The scaling behavior observed near critical fields $H_{1}^{\ast}$ and $H_{2}^{\ast}$ is consistent with a model where superconductivity is destroyed by quantum phase fluctuations in a 2D superconductor.  We note that such scaling
is independent of the nature of the insulator\cite{MPAF-SIT,Sachdev-book}.  In particular, the same scaling, except for the value of $z\nu$, is expected when the insulating phase is due to disorder and when it is caused by inhomogeneous charge ordering, which is known to be present in low-doped LSCO (refs. \onlinecite{Julien99,Singer02NQR,Dumm03EM,Ando02Ranisotropy,Ando03MR}), including cluster charge glass\cite{Ivana-PRL,Ivana-PRB2011,XShi-NatMat} and charge density wave\cite{Lake}.  

We expect a similar two-stage $H$-field-tuned SIT to occur for other doping levels in the entire underdoped superconducting regime.  This is supported by our observation of the same behavior in samples that were prepared in two very different ways and in which the number of doped holes was probably not exactly the same.  Also, the possibility of two critical points was suggested recently\cite{Leridon} for LSCO films with $x=0.08, 0.09$ and 0.10 (Supplementary Information).  
Our results provide important insight into the interplay of vortex line physics and quantum criticality.
Apparently, the high-temperature, clean-limit SIT quantum critical fluid falls into the grip of disorder upon lowering $T$ and splits into two transitions: first, a $T=0$ vortex lattice to vortex glass transition, followed by a genuine superconductor-insulator QPT at higher $H$.  

\begin{methods}
\noindent {\bf Samples.}  The LSCO film with a nominal $x=0.07$ was described in detail in ref. \onlinecite{XShi-NatMat}.  The LSCO single crystal with a nominal $x=0.06$ was grown by the traveling-solvent floating-zone technique\cite{SasagawaLSCO}.  Measurements were carried out on a sample that was cut out along the main crystallographic axes and polished into a bar with dimensions $3.02\times 0.42\times 0.34$~mm$^3$ suitable for direct measurements of the in-plane resistance.  Electrical contacts were made by evaporating gold on polished crystal surfaces, followed by annealing in air at $700^{\circ}$C. For current contacts, the whole area of two opposing side faces was covered with gold to ensure a uniform current flow through the sample.  In turn, the voltage contacts were made narrow ($\sim80~\mu$m) in order to minimize the uncertainty in the absolute values of the resistance.  The distance between the voltage contacts is 1.41~mm.  Gold leads were attached to the sample using the Dupont 6838 silver paste. This was followed by the heat treatment at $450^{\circ}$C in the flow of oxygen for 6 minutes.  The resulting contact resistances were less than 1~$\Omega$ at room $T$.  As a result of annealing, the sample composition is probably La$_{1.94}$Sr$_{0.06}$CuO$_{4+y}$, with $y$ not precisely known.
\vspace*{12pt}

\noindent{\bf Measurements.}  The in-plane sample resistance and magnetoresistance were measured with a standard four-probe ac method ($\sim11$~Hz) in the Ohmic regime at current densities as low as  $(3\times 10^{-3}-3\times 10^{-2})$~A/cm$^2$ for the $x=0.07$ film and $7\times 10^{-2}$~A/cm$^2$ for the $x=0.06$ crystal.  To cover a wide range of $T$ and $H$, several different cryostats were used:  a $^3$He system at $T$ down to 0.3~K and with magnetic fields up to 9~T, using $0.02-0.05$~T/min sweep rates; a dilution refrigerator with $T$ down to 0.02~K and a $^3$He system (0.3~K$\leq T\lesssim$30~K) in superconducting magnets with fields up to 18~T at the National High Magnetic Field Laboratory (NHMFL), using $0.1-0.2$~T/min sweep rates; and a 35~T resistive magnet at the NHMFL with a variable-temperature insert (1.2~K$\leq T\lesssim$30~K), using 1~T/min sweep rates.  Therefore, the
MR measurements span more than two orders of magnitude in $T$, down to 0.09~K, \textit{i.e.} much lower than temperatures commonly employed in studies of underdoped cuprates.  For that reason, we use excitations (\textit{i.e}. current densities) that are orders of magnitude lower than in similar studies.  
It was not possible to cool the samples below 0.09~K.  The fields, applied perpendicular to the CuO$_2$ planes, were swept at constant temperatures. The sweep rates were always low enough to avoid the heating of the sample due to eddy currents.

The resistance per square $R_{\square}=\rho/d=\rho/(nl)$ ($d$ -- sample thickness, $n$ -- number of CuO$_2$ layers, $l=6.6$~\AA\, -- thickness of each layer); the resistance per square per CuO$_2$ layer $R_{\square/layer}=nR_{\square}=\rho/l$.

\end{methods}

\pagebreak

\noindent\textbf{\large{Acknowledgements}}

\noindent We thank P. Baity for experimental assistance, and A.T. Bollinger and I. Bo\v{z}ovi\'c for the film sample.  The work by X.S., P.V.L. and D.P. was supported by NSF/DMR-0905843, NSF/DMR-1307075, and the NHMFL, which was supported by NSF/DMR-0654118, NSF/DMR-1157490 and the State of Florida.  The work of T.S. was supported by MSL Collaborative Research Project.  V.D. was supported by NSF/DMR-1005751.  V.D. and D.P. thank the Aspen Center for Physics, where part of this work was done, for hospitality and support under NSF/PHYS-1066293.

\noindent\textbf{\large{Author contributions}}

\noindent X.S. and D.P. conceived the project; the single crystal was grown by T.S.; X.S. and P.V.L performed the measurements and analysed the data; V.D. and D.P. contributed to the data analysis and interpretation; X.S., P.V.L and D.P. wrote the manuscript; D.P. planned and supervised the investigation.  All authors commented on the manuscript.

\noindent\textbf{\large{Additional information}}

\noindent The authors declare that they have no competing financial interests.  Supplementary information accompanies this paper.  Correspondence and requests for materials should be addressed to D.P.~(email: dragana@magnet.fsu.edu).

\newpage

\noindent\textbf{\Large{Supplementary Information}}

\makeatletter
\makeatletter \renewcommand{\fnum@figure}{{\bf{\figurename~S\thefigure}}}
\makeatother

\setcounter{figure}{0}

\noindent\textbf{\large{{Characteristic temperatures ${\bm{T_c(H)}}$ and ${\bm{T_{max}(H)}}$}}}

\begin{figure}[h]
\centerline{\includegraphics[width=10cm]{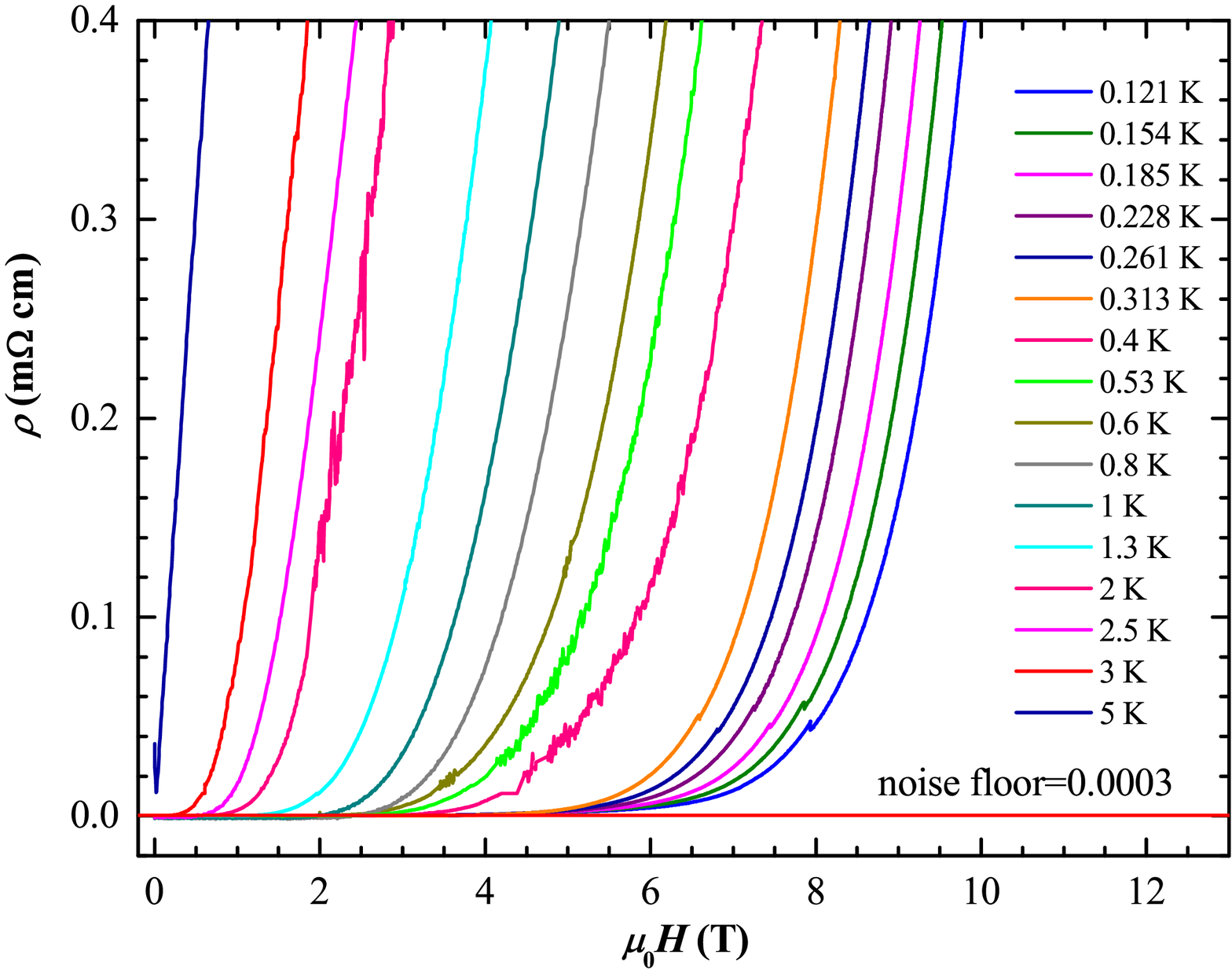}}
  \caption{\textbf{Determination of the zero-resistance ${\bm{T_c}}$ in ${\bm{x=0.07}}$ LSCO film.}  In a magnified plot of $\rho(H)$ curves at fixed temperatures, the field where the resistivity is smaller than the experimental noise floor (here $3\times 10^{-4}$~m$\Omega\cdot$cm) is defined as the ``zero-resistance field'' at a given temperature.  That temperature, in turn, is identified as the zero-resistance $T_{c}(H)$ in this field.}
  \label{fig:zeroRTc}
\end{figure}

\begin{figure}
\centerline{\normalsize{\textbf a}\includegraphics[width=8cm]{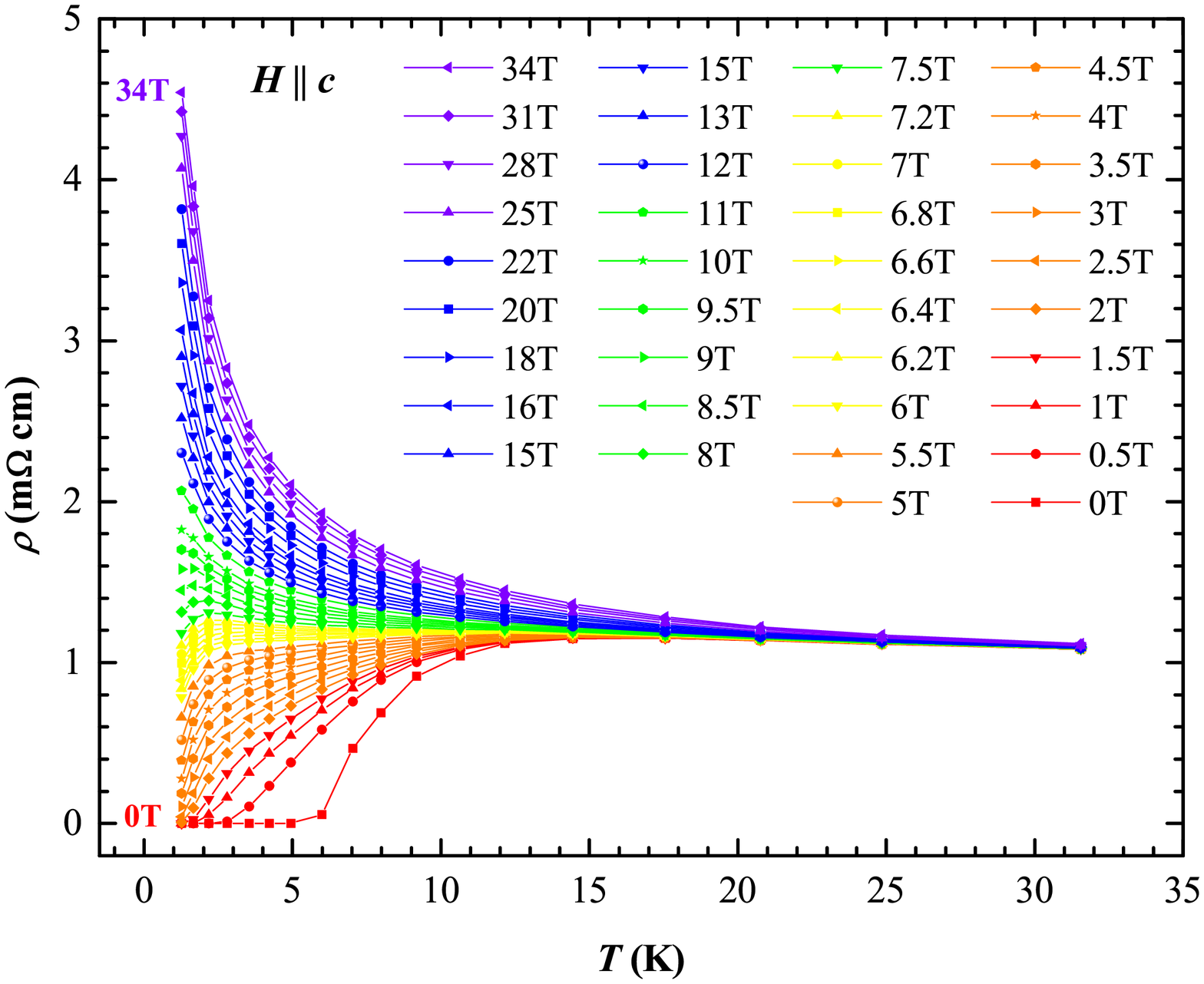}
\normalsize{\textbf b}\includegraphics[width=8cm]{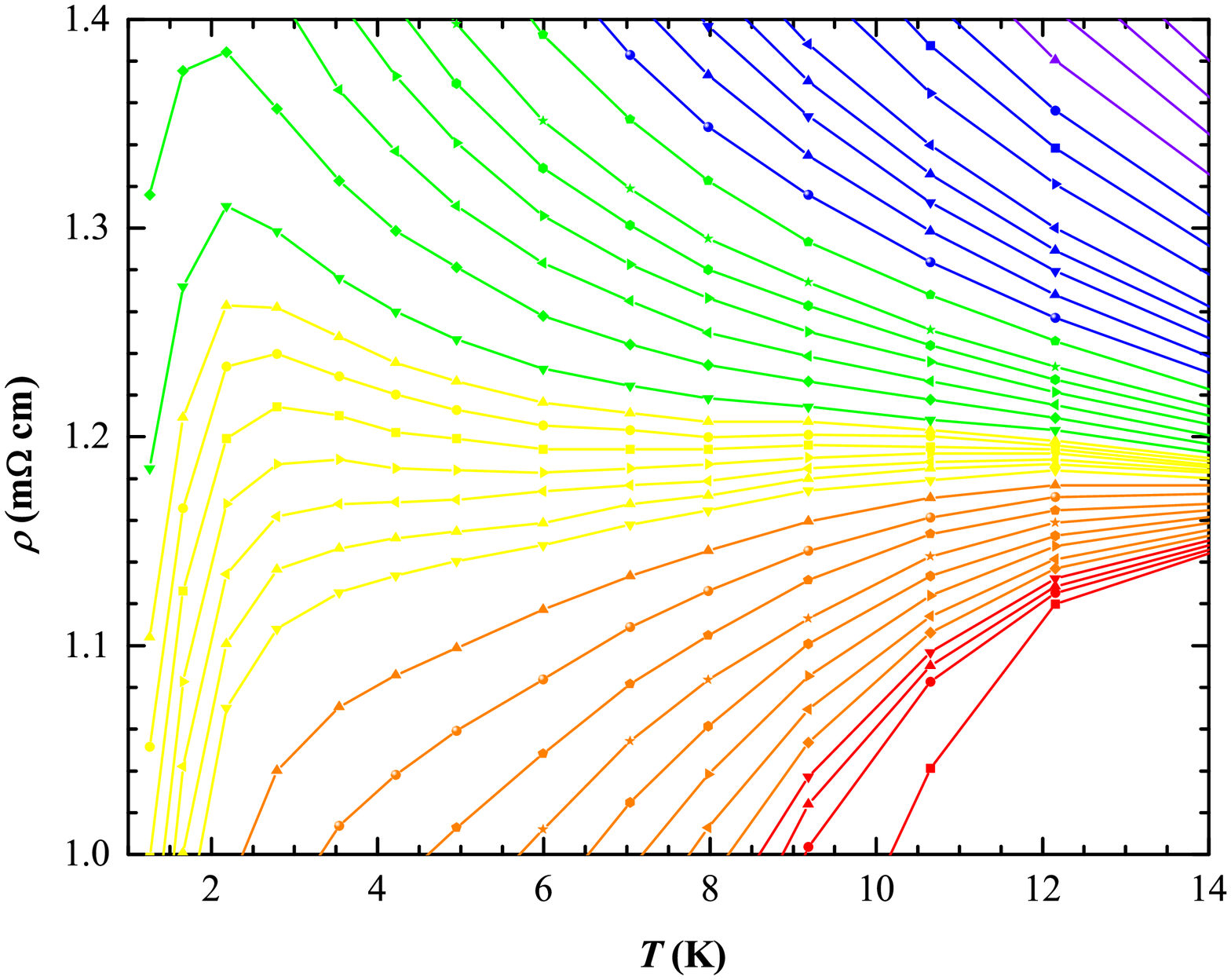}}
  \caption{\textbf{Temperature dependence of the in-plane resistivity ${\bm{\rho}}$ in different magnetic fields ${\bm{H\parallel c}}$ for ${\bm{x=0.06}}$ LSCO single crystal.  a,} $\rho(T,H)$ data are shown for 1.2~K$< T<$35~K and $H$ up to 34~T.  \textbf{b,} $\rho(T,H)$ in \textbf{a} shown magnified in the intermediate field regime, where a maximum in $\rho(T)$ emerges at low temperatures.}
  \label{fig:RvsT-6per}
\end{figure}

\begin{figure}
\centerline{\normalsize{\textbf a}\includegraphics[width=10cm]{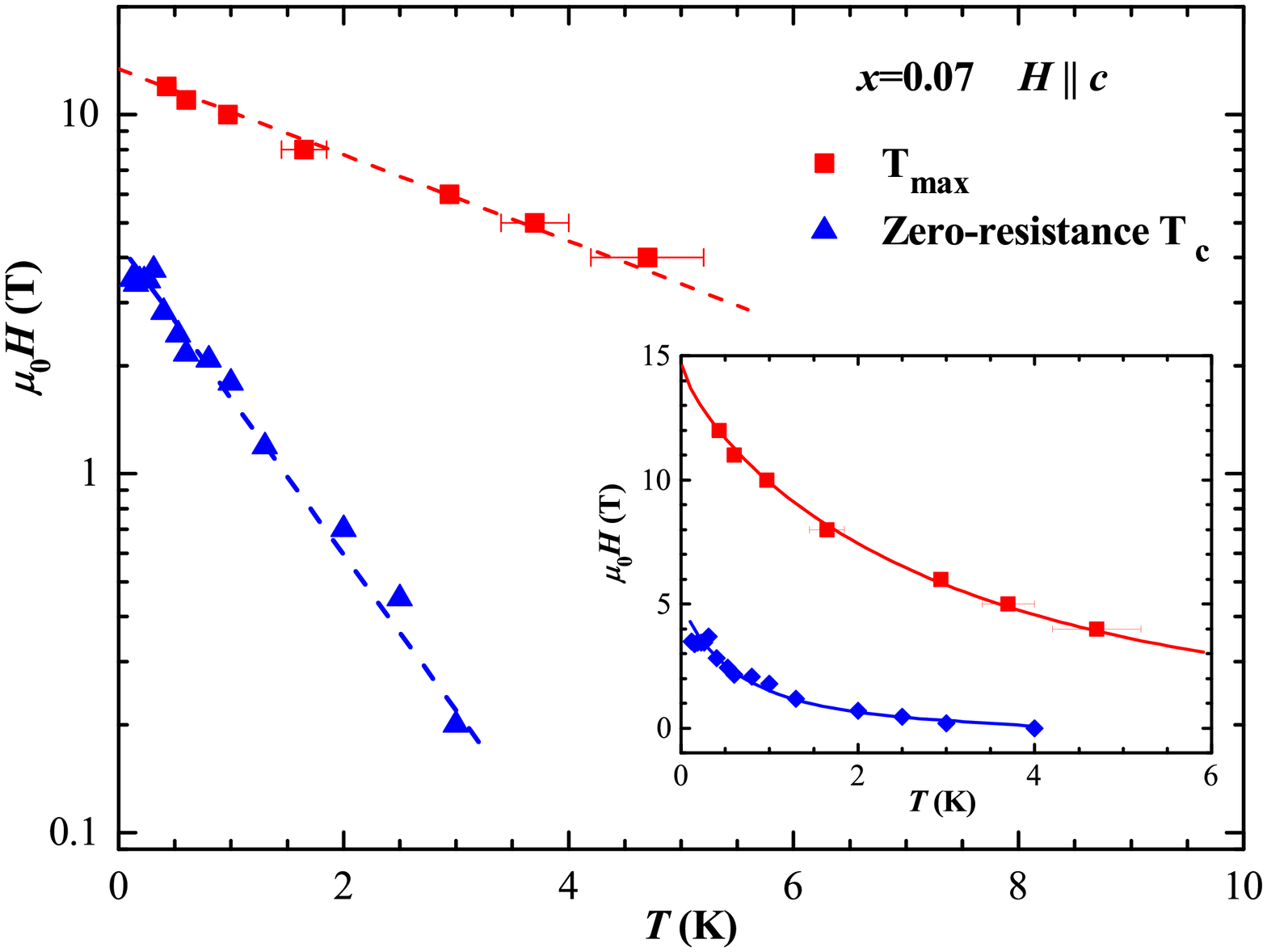}}\vspace*{0.5cm}
\centerline{\normalsize{\textbf b}\includegraphics[width=10cm]{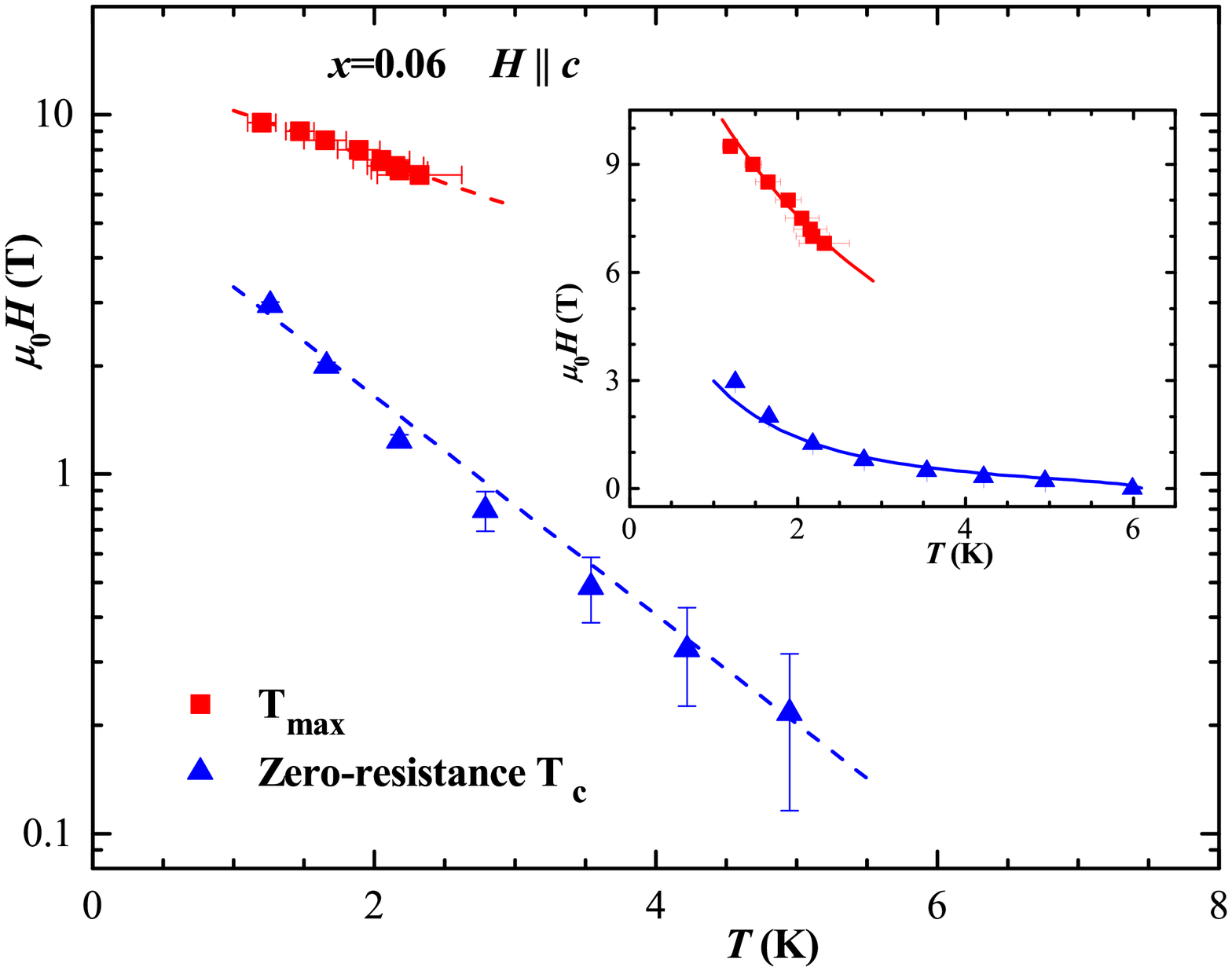}}
  \caption{\textbf{${\bm {T_c(H)}}$ and ${\bm {T_{max}(H)}}$ for two different samples.}  The dashed lines show phenomenological exponential fits $H(T)=H_0\exp(-T/T_0)$ for $x=0.07$ LSCO film in \textbf{a}, and $x=0.06$ single crystal in \textbf{b}.  Solid lines in the insets show fits to the same data using the expression for the thermal melting of the vortex lattice (Eq.~\ref{eqmelting}).}
  \label{fig:Tcfits}
\end{figure}

The zero-resistance $T_c(H)$ was determined as illustrated in Fig.~S\ref{fig:zeroRTc}.  $T_{max}(H)$ is the temperature at the maximum of the $\rho(T)$ curves in the intermediate-field regime (see Figs.~1a and S\ref{fig:RvsT-6per}).  Figure~S\ref{fig:Tcfits} shows $T_c(H)$ and $T_{max}(H)$ with the phenomenological fits $H(T)=H_0\exp(-T/T_0)$ (or, equivalently, $T_c(H)=T_0\ln(H_0/H)$ for $H>0$).  The following fitting parameters are obtained.  For $T_c(H)$: $\mu_{0}H_0=(4.4\pm 0.5)$~T and $T_0=(1.0\pm 0.2)$~K for $x=0.07$ film (Fig.~S\ref{fig:Tcfits}a), $\mu_{0}H_0=(6.2\pm 0.3)$~T and $T_0=(1.4\pm 0.1)$~K for $x=0.06$ single crystal (Fig.~S\ref{fig:Tcfits}b).  For $T_{max}(H)$: $\mu_{0}H_0=(13.4\pm 0.7)$~T and $T_0=(3.6\pm 0.3)$~K for $x=0.07$ film, $\mu_{0}H_0=(14.0\pm 0.6)$~T and $T_0=(3.2\pm 0.1)$~K for $x=0.06$ single crystal.  A similar exponential decay was found in the irreversibility field $H_{irr}(T)$ in magnetometry measurements in LSCO [S1], with $T_0\sim 1$~K in $x=0.06$ single crystal.  The corresponding irreversibility temperature in a given field, $T_{irr}(H)$, is identified from the resistivity measurements as the zero-resistance $T_c(H)$ [S2, S3].  

We note that, in general, $T_{irr}(H)$ (\textit{i.e.} $T_c(H)$) is distinct from the thermal melting transition of the vortex solid to a vortex liquid at $T_m(H)>T_{irr}(H)$ (see sketch in Fig.~5) [S4].  Nevertheless, the zero-resistance $T_c(H)$ data are sometimes fitted to the melting curve given by [S5, S6]
\begin{equation}
\frac{\sqrt{b_m(t)}}{1-b_m(t)}\frac{t}{\sqrt{1-t}}\left[ \frac{4 \left(\sqrt{2} - 1 \right)}{\sqrt{1-b_m(t)}} + 1 \right]
= \frac{2 \pi c_L^2}{\sqrt{Gi}},
\label{eqmelting}
\end{equation}
where $t \equiv T/T_c$, $b_m(t) \equiv B_m/\mu_0 H_{c2}(t)=B_m/[\mu_{0}H_{c2}(0)(1-t)]$, $c_L$ is the Lindemann number, and $Gi$ is the Ginzburg number.  The Lindemann number $c_L$ represents the ratio of the root-mean-square amplitude of vortex lattice thermal fluctuations over the lattice constant.  The Ginzburg number $Gi$ is given by
\begin{align}
Gi &= \frac{1}{2} \left( \frac{k_B T_c \gamma}{\frac{4 \pi}{\mu_0} \left(\mu_0 H_c(T=0)\right)^2 \xi^3_0}\right)^2 \label{eqgi} \\
     &\approx \left(9.225 \times 10^{8} \left[\mathrm{Wb}^{-1}\mathrm{K}^{-1}\right] \times \mu_0 H_{c2}(0) T_c \lambda_{ab} \lambda_c\right)^2,
\label{eqgi1}
\end{align}
where $\gamma$ is the anisotropy ratio $\gamma \equiv \lambda_{c}/\lambda_{ab}$, and the definition $H_{c2}(0) \equiv \frac{4 \pi}{\mu_0} \lambda_{ab}^2 \left[\mu_0 H_c(T\!=\!0)\right]^2/\Phi_0$ has been used ($\lambda_{ab}$ and $\lambda_{c}$ are the penetration depths parallel and perpendicular to the $ab$-plane at zero temperature; $\Phi_0$ is the flux quantum, and the mean-field Ginzburg-Landau coherence length $\xi_0$ can be represented by $H_{c2}(0)$ from $\mu_0 H_{c2}(T=0)=\Phi_0/(2\pi\xi_0^2)$).  The right hand side of Eq.~\ref{eqmelting} can be thus rewritten as
\begin{equation}
\frac{2 \pi c_L^2}{\sqrt{Gi}} \approx \frac{K}{\mu_{0}H_{c2}(0) T_c},
\label{eqconstK}
\end{equation}
where $K \equiv 2 \pi c_L^2/\left(9.225 \times 10^{8} \left[\mathrm{Wb}^{-1}\mathrm{K}^{-1}\right] \times  \lambda_{ab} \lambda_c\right)$ is a fitting parameter in addition to $H_{c2}(0)$ and $T_c$.  Figure~S\ref{fig:Tcfits} insets show the melting line fits to our $T_c(H)$ data with the following fitting parameters:
$\mu_{0}H_{c2}(0)=(5\pm 1)$~T, $T_c=(4\pm 1)$~K and $K=(14\pm 5)$~T$\cdot$K for $x=0.07$ film (Fig.~S\ref{fig:Tcfits}a inset), $\mu_{0}H_{c2}(0)=(9\pm 2)$~T,  $T_c=(6.1\pm 0.5)$~K and $K=(27\pm 5)$~T$\cdot$K for $x=0.06$ single crystal (Fig.~S\ref{fig:Tcfits}b inset).  The values of $K$ obtained from the fits are consistent with  $\lambda_{ab}\sim 1~\mu$m and $\gamma\sim 10$ for LSCO, and $c_L\sim 0.1-0.4$ [S4, S6-S8].  

Both the phenomenological exponential fit and the melting line fit (Eq.~\ref{eqmelting}) describe the $T_c(H)$ data reasonably well.  In fact, even the $T_{max}(H)$ values can be fitted well using  Eq.~\ref{eqmelting} with the following fitting parameters: $\mu_{0}H_{c2}(0)=(15\pm 2)$~T, $T_c=(25\pm 5)$~K and $K=(160\pm 20)$~T$\cdot$K for $x=0.07$ film (Fig.~S\ref{fig:Tcfits}a inset), $\mu_{0}H_{c2}(0)=(17\pm 3)$~T, $T_c=(14\pm 2)$~K and $K=(160\pm 20)$~T$\cdot$K for $x=0.06$ single crystal (Fig.~S\ref{fig:Tcfits}b inset), where the obtained values of $K$ are still reasonable.  However, there is no known reason to associate $T_{max}(H)$ with the thermal melting of the vortex solid.  Likewise, as noted above, $T_c(H)$ is, in general, distinct from the thermal melting transition [S4].  In addition, since the melting line fits have much larger error bars than the phenomenological exponential fits in all cases, the $H_0$ values obtained from exponential fits are used in the discussions in the main text.  The precise values of $H_0$, however, do not affect any of our conclusions.

\noindent\textbf{\large{{Scaling near the SIT}}}

Scaling arguments [S9] imply that the critical resistance (\textit{i.e.} $R_{\square/layer}$ at the SIT) for a given type of system must be universal.  We do find that the critical resistances at $H_{1}^{\ast}$ are almost the same in both samples and estimate that they would be of the same order of magnitude at $H_{2}^{\ast}$.  The precise value of the critical resistance for the SIT, however, may depend on the nature of interactions and disorder, and its understanding remains an open question [S10, S11].

The possibility of two critical points was suggested recently [S12] for LSCO films with $x=0.08, 0.09$ and 0.10.  
However, unlike our study that was performed down to $0.09$~K, those measurements extended down to only 1.5~K making it impossible to identify the form of $\rho(T)$ for $T<T_{max}$.  In addition, $T_c(H)$ and $T_{max}(H)$ were not discussed and thus no connection was made to $H_{1}^{\ast}$ and $H_{2}^{\ast}$, respectively.  The scaling regions associated with the two critical points were also not identified, $H_{c}'$ and $\Delta\sigma_{SCF}$ were not determined, and the phase diagram was not constructed.

\begin{figure}
\centerline{\normalsize{\textbf a}\includegraphics[width=10cm]{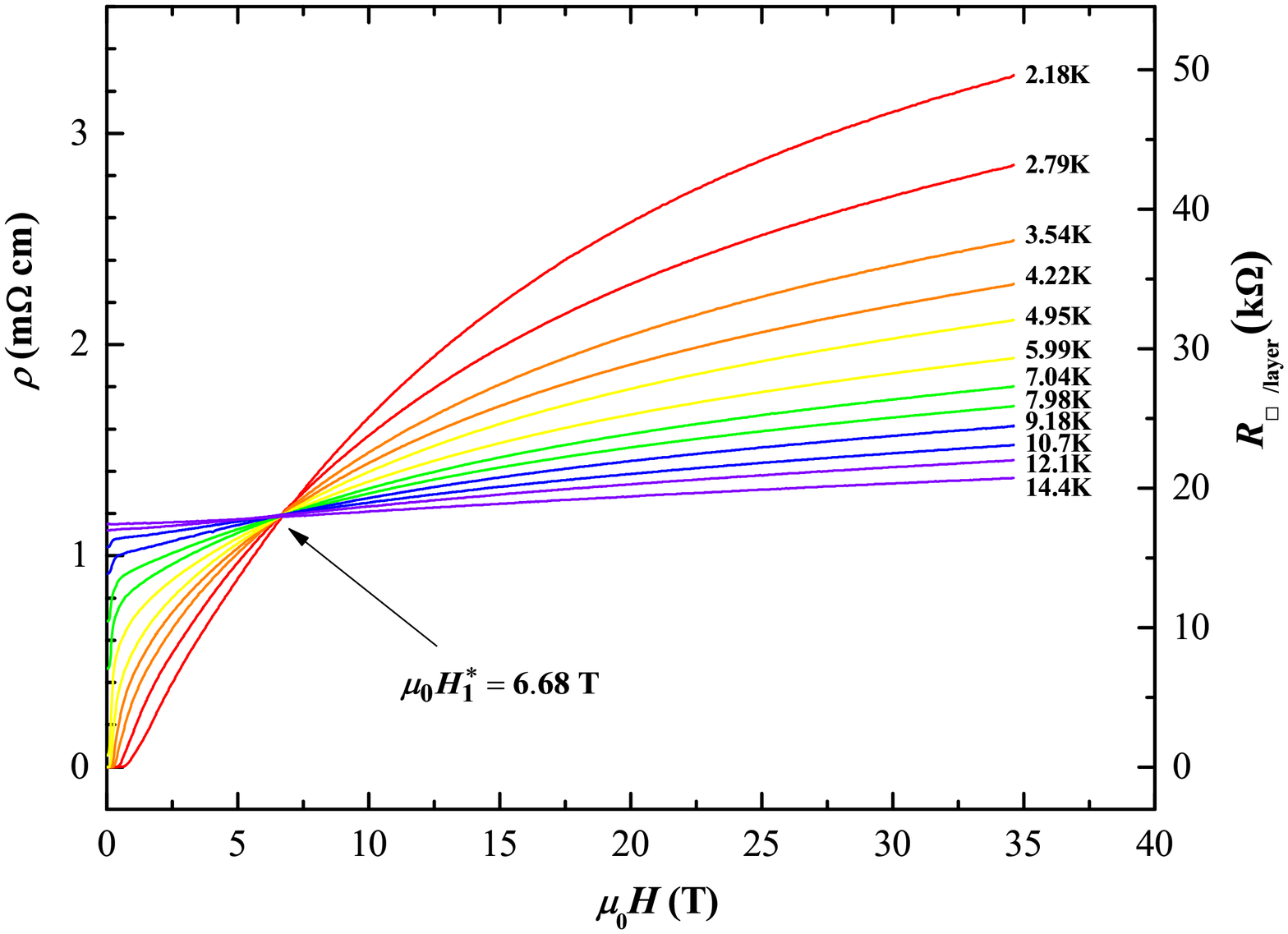}}\vspace*{0.5cm}
\begin{minipage}[t]{0.5\textwidth} \normalsize{\textbf b}\includegraphics[width=8.6cm]{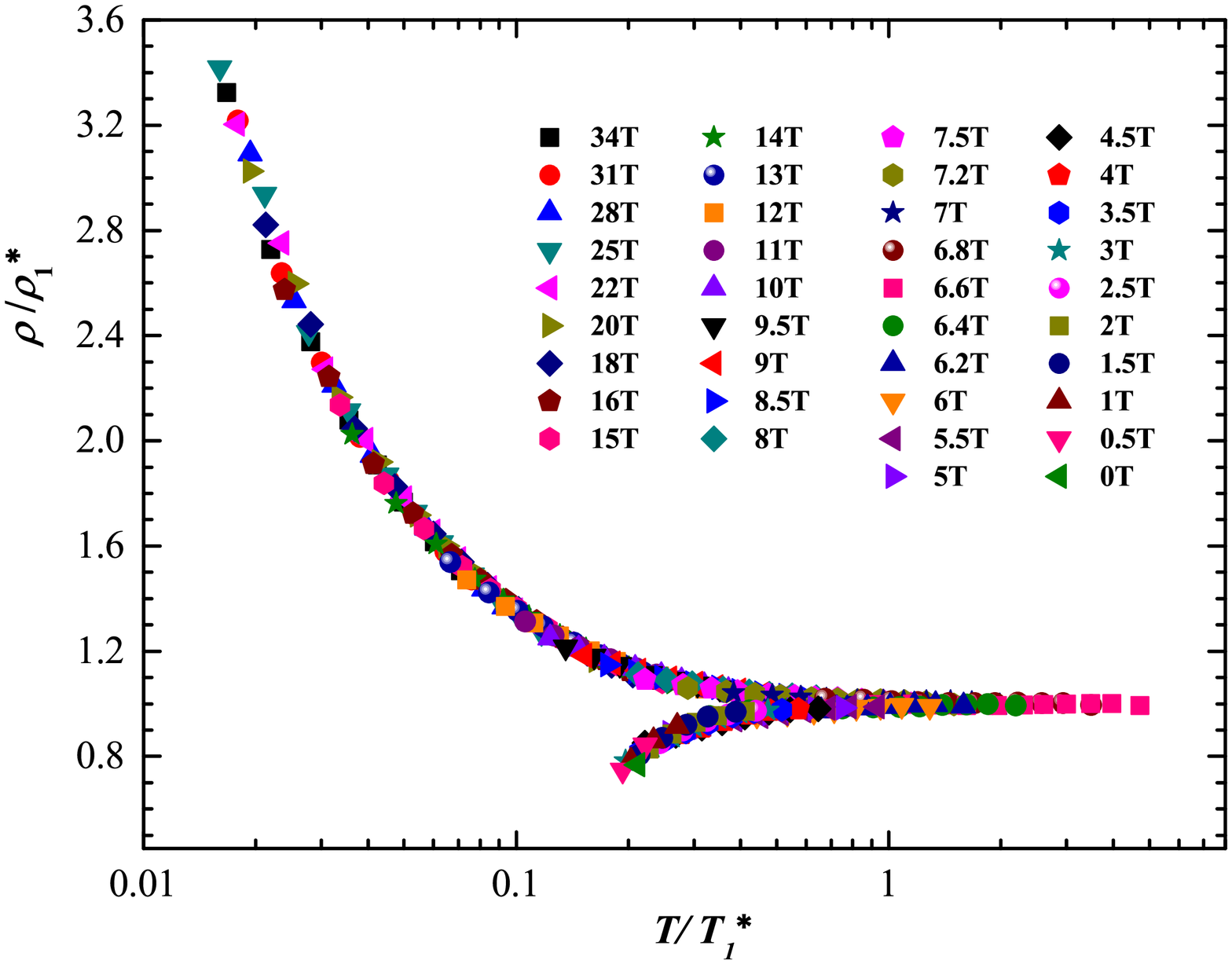}
\end{minipage}
\begin{minipage}[t]{0.5\textwidth} \hspace*{0.5cm}\normalsize{\textbf c}\includegraphics[width=8.5cm]{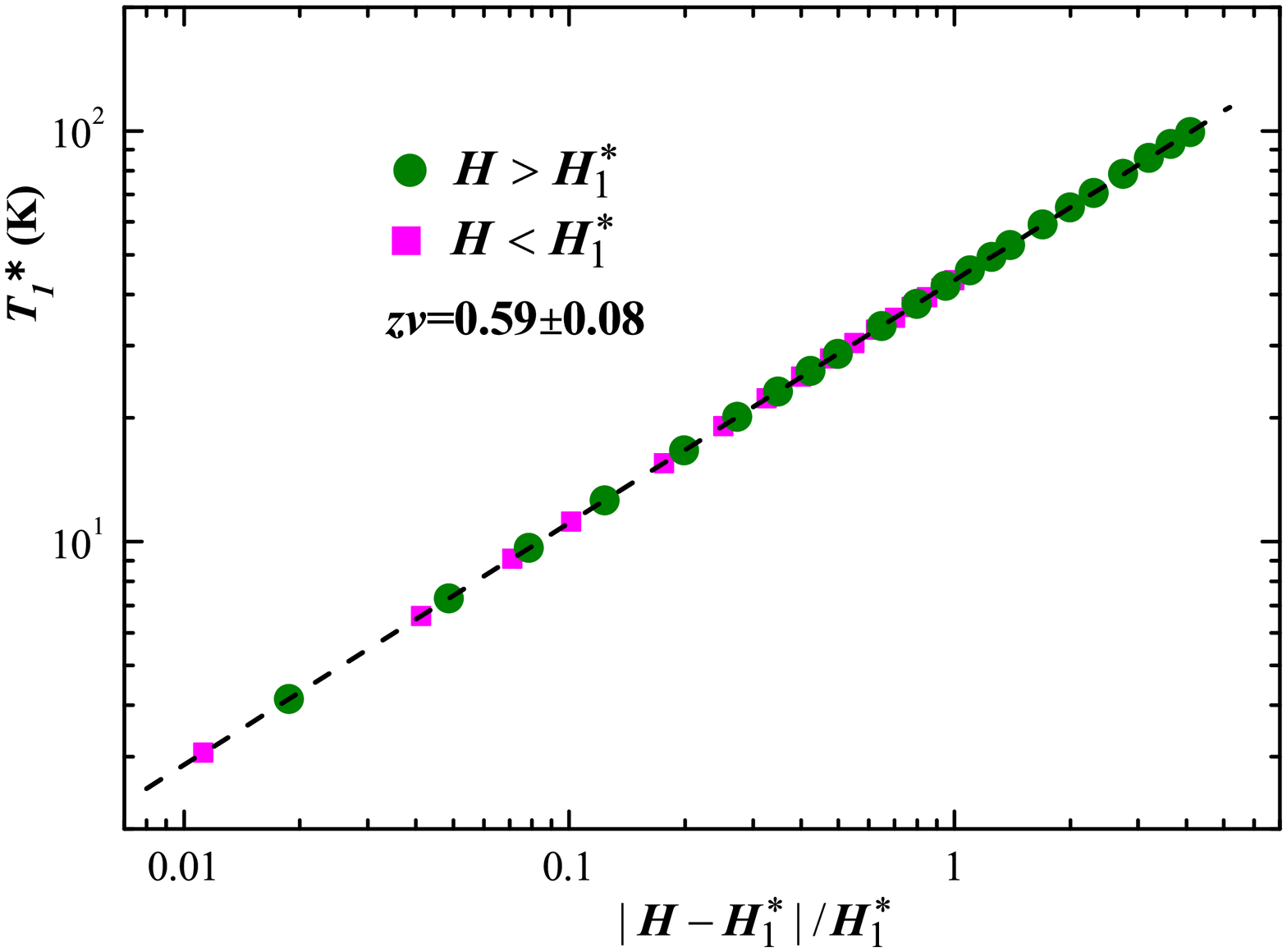}
\end{minipage}
\caption{\textbf{High-temperature ($T\gtrsim 2$~K) behavior of the resistivity ${\bm{\rho}}$ in different magnetic fields ${\bm{H\parallel c}}$ for ${\bm{x=0.06}}$ LSCO single crystal.}
 \textbf{a,} Isothermal $\rho(H)$ curves in the high-$T$ region show the existence of a $T$-independent crossing point at $\mu_{0}H_{1}^{\ast}=6.68$~T and $\rho_{1}^{\ast}=1.19$~m$\Omega\cdot$cm (or $R_{\square/layer}\approx 18.0$~k$\Omega$).  \textbf{b,} Scaling of the data in \textbf{a} with respect to a single variable $T/T_{1}^{\ast}$.  The scaling region is shown in more detail in Fig.~S\ref{fig:crystalphase}.  
The upper branch of the scaling curve is made of the $H>H_{1}^{\ast}$ data, while the lower branch corresponds to $H<H_{1}^{\ast}$.  \textbf{c,} The scaling parameter $T_{1}^{\ast}$ as a function of $\mid H-H_{1}^{\ast}\mid/H_{1}^{\ast}$ on both sides of $H_{1}^{\ast}$.  The dashed line is a fit with slope $z\nu\approx 0.59$.}
  \label{fig:6perscaling}
\end{figure}

\noindent\textbf{\large{{Superconducting fluctuations}}}

\begin{figure}[h]
\vspace*{24pt}
\centerline{\includegraphics[width=10cm]{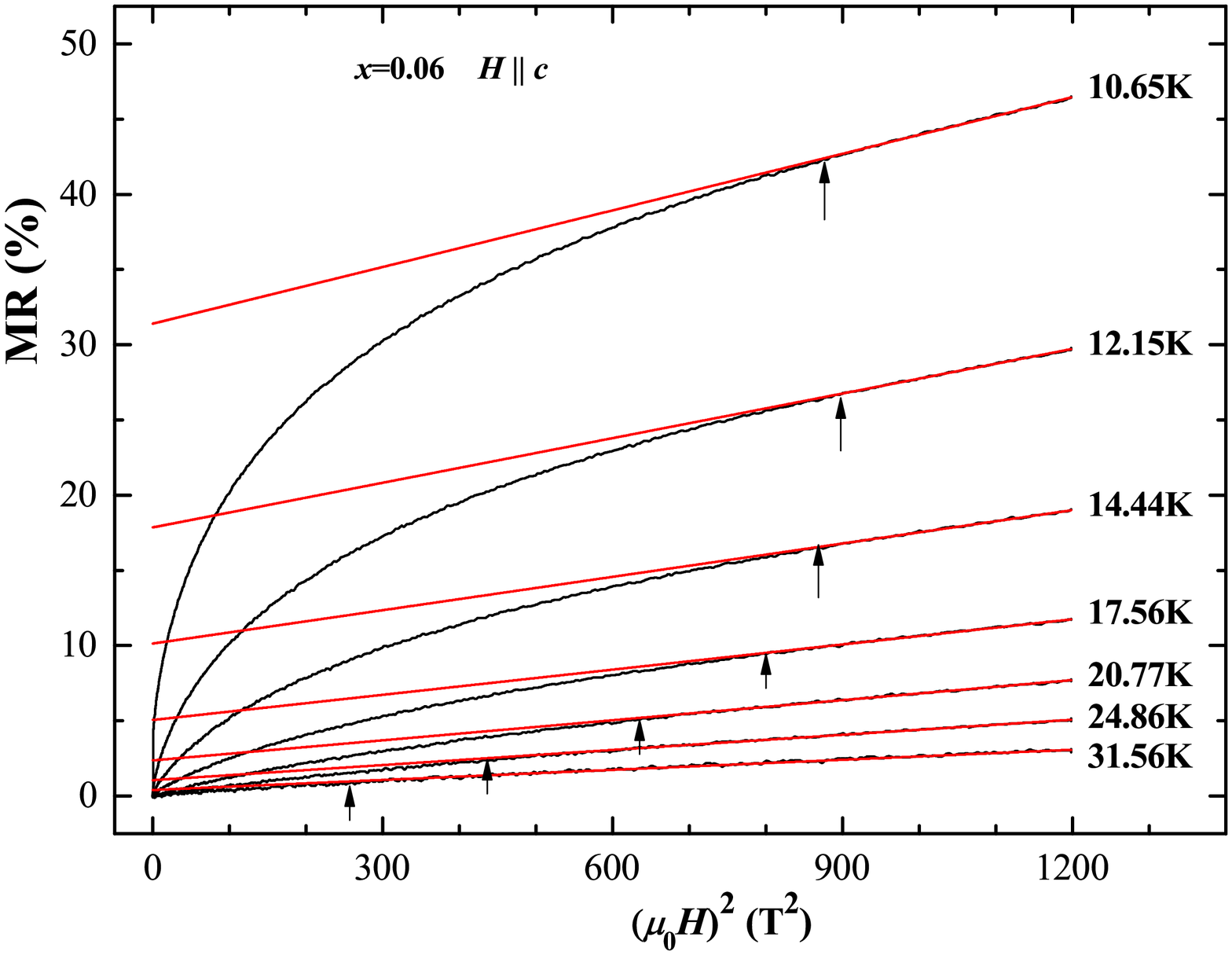}}
  \caption{\textbf{Transverse (${\bm{H\parallel c}}$) in-plane magnetoresistance \textit{vs.} ${\bm{H^2}}$ for the ${\bm{x=0.06}}$ single crystal.}  Red lines are fits representing the contributions from normal state transport, i.e. they correspond to $[R(H)-R(0)]/R(0) = [R_n(0)-R(0)]/R(0) +\alpha H^2$. The intercept of the red line shows the relative difference between the fitted normal state resistance and the measured resistance at zero field.  The difference between the red lines and the measured magnetoresistance is due to the superconducting contribution.  Arrows show $H_c'$, the field above which superconducting fluctuations are fully suppressed and the normal state is restored.}
  \label{fig:MRH2}
\end{figure}

\begin{figure}
\centerline{\includegraphics[width=10cm]{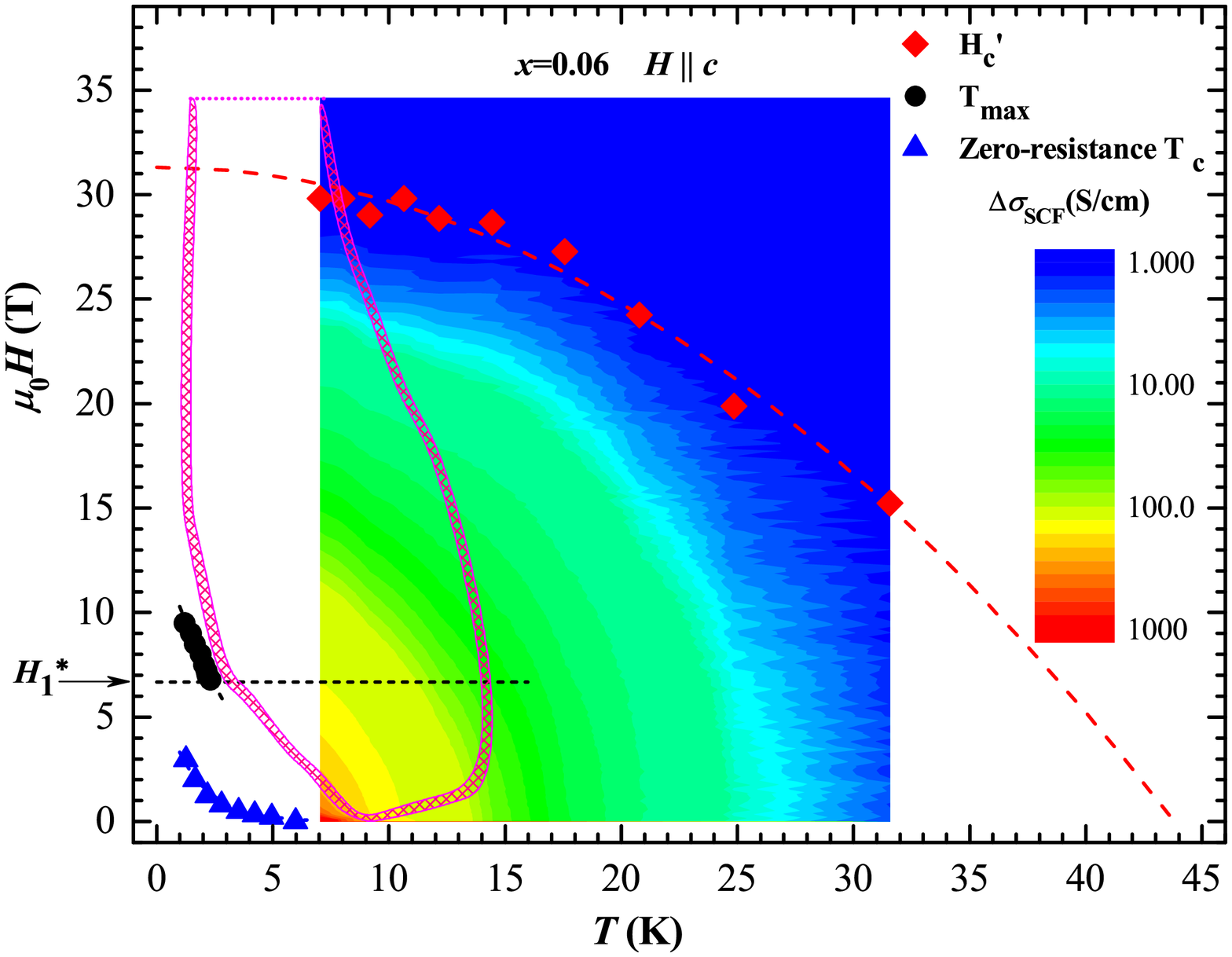}}  
\caption{\textbf{Transport ${\bm{H-T}}$ phase diagram and scaling region for the ${\bm{x=0.06}}$ single crystal.}  The color map (on a log scale) shows the SCF contribution to conductivity $\Delta\sigma_{SCF}$ as a function of $T$ and $H\parallel c$.  The dashed red line is a fit with $\mu_{0}H_{c}^{'}$[T]$=31[1-(T$[K]$/44)^{2}]$.  The dashed black line indicates the value of the $T=0$ critical field $H_{1}^{\ast}$ for scaling.  The pink lines show the high-$T$ scaling region: the hashed symbols mark areas beyond which scaling fails, and the dots indicate areas beyond which the data are not available.  Even though this sample was measured only down to 1.2~K and thus the low-T scaling and $H_{2}^{\ast}$ were not observed, the shape of its high-$T$ scaling region closely resembles that of the $x=0.07$ film sample (Fig.~4a).  In particular, it is consistent with the presence of another transition near 14~T, where $T_{max}(H)$ extrapolates to zero.}
  \label{fig:crystalphase}
\end{figure}

For both samples, the field $H_{c}'(T)$, determined as shown in Fig.~S\ref{fig:MRH2}, is fitted by $H_{c}'(T)=H_{c}'(0)[1-(T/T_2)^2]$ (see Fig.~S\ref{fig:crystalphase} for the $x=0.06$ single crystal).
Although the values of $T_c(H=0)$ in these samples are very similar, their $H=0$ onset temperatures for SCFs, $T_2$, are very different.  In particular, 
$T_2=29$~K in the film is consistent with the results from terahertz spectroscopy [S13] obtained on similar films.  This value is lower than $T_2=44$~K found in the single crystal, but the 
latter is in agreement with those determined from the onset of diamagnetism [S14] and the Nernst effect [S15] in LSCO crystals.  Likewise, $\mu_{0}H_{c}'(T=0)=31$~T in the crystal is a factor of two larger than $\mu_{0}H_{c}'(T=0)=15$~T in the film.  These observations suggest that the origin of the discrepancy between onset temperatures for SCFs and $H_{c}'(T=0)$ obtained from different experiments may be, at least partly, attributed to the differences in the sample preparation conditions.  

We remark that $H_{c}'(T)$ values determined from the MR measurements have been shown to decrease with underdoping [S16], including also non-superconducting samples [S17].  

\begin{figure}
\centerline{\normalsize{\textbf a}\includegraphics[width=10cm]{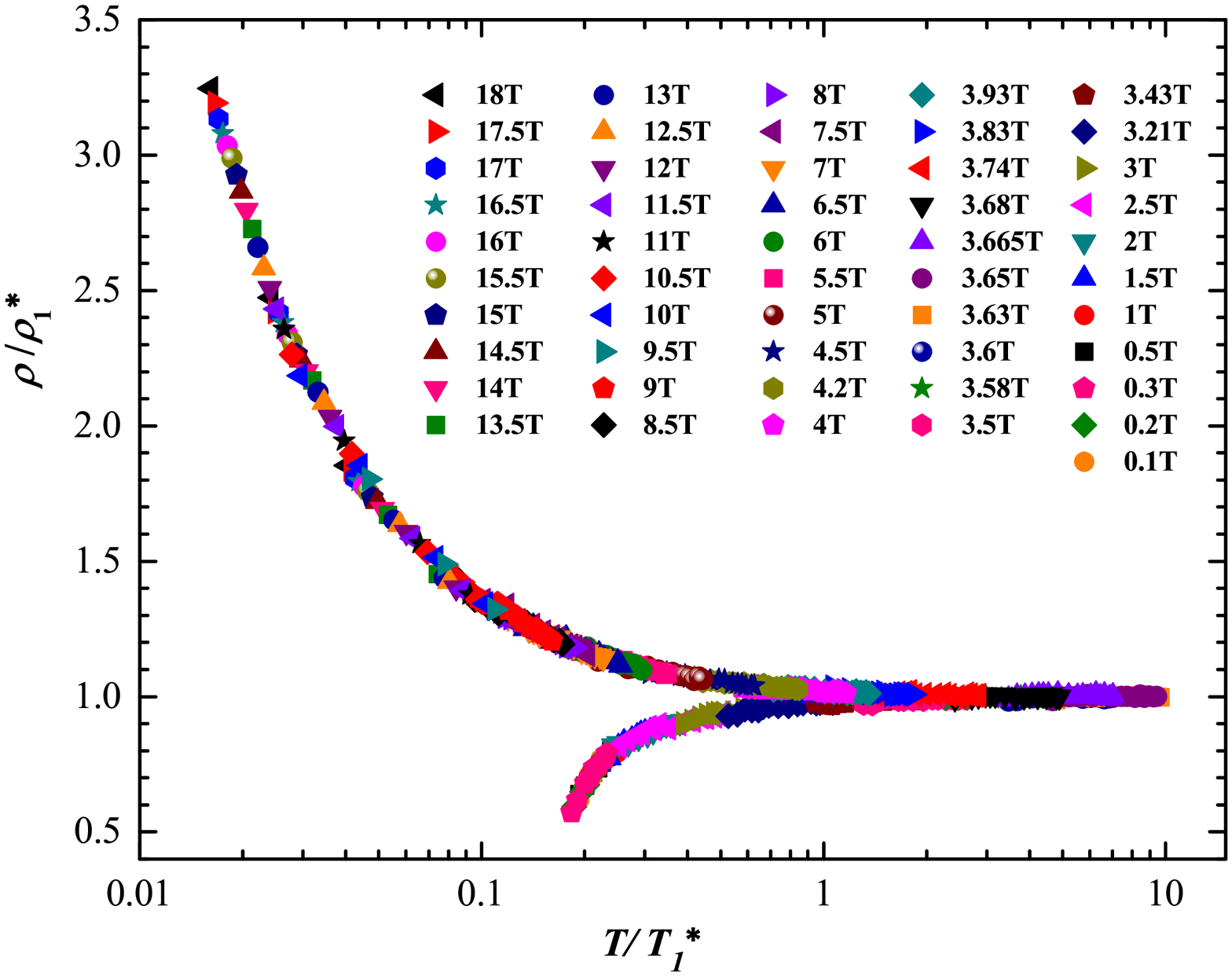}}\vspace*{0.5cm}
\centerline{\normalsize{\textbf b}\includegraphics[width=10cm]{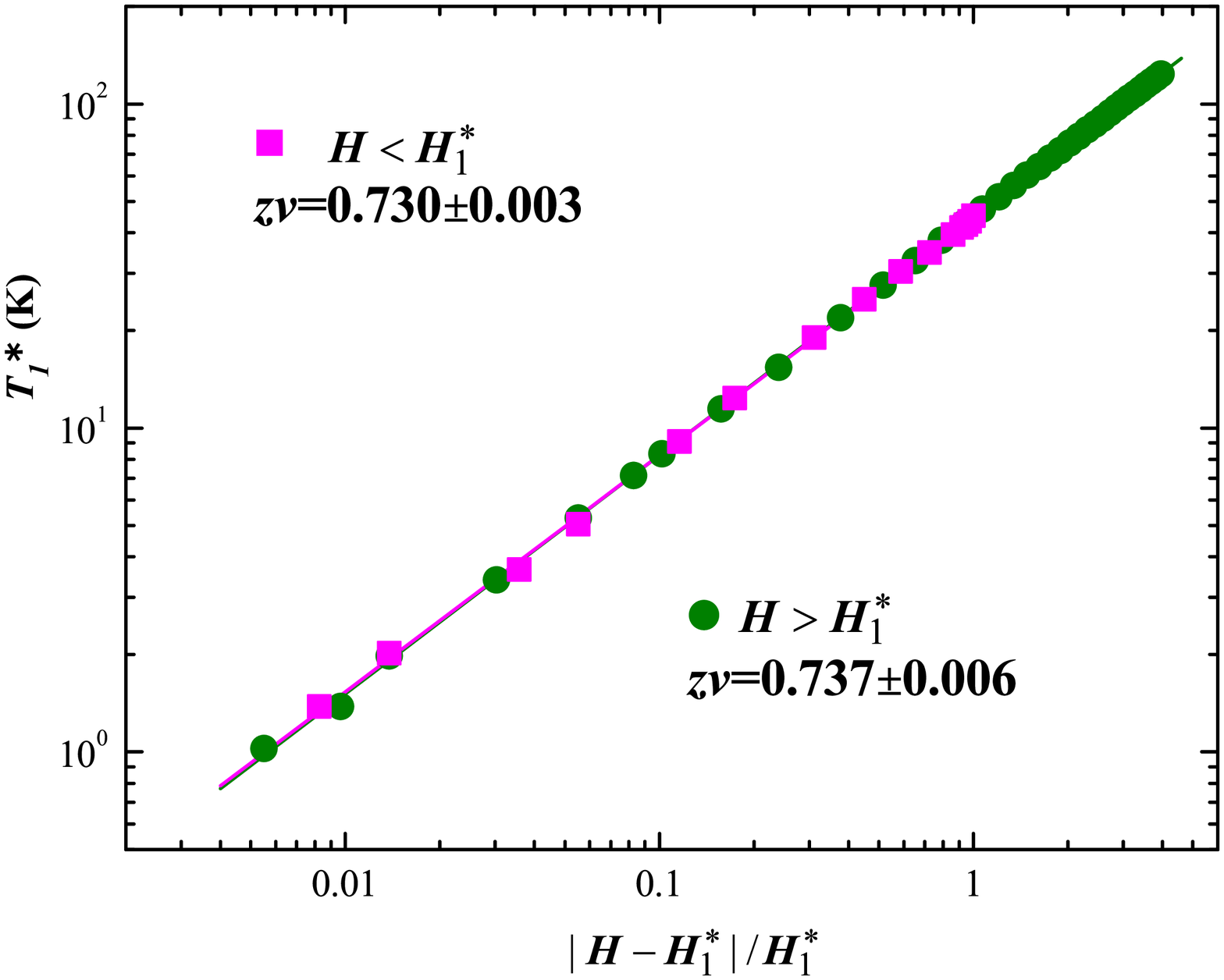}}
  \caption{\textbf{Scaling of ${\bm{\rho(T,H)}}$ near ${\bm{\mu_{0}H_{1}^{\ast}=3.63}}$~T for ${\bm{x=0.07}}$ LSCO film.}  \textbf{a,} Scaling of $\rho(T,H)$ with respect to a single variable $T/T_{1}^{\ast}$ over the entire scaling region shown in Fig.~4a.  \textbf{b,} The scaling parameter $T_{1}^{\ast}$ as a function of $\mid H-H_{1}^{\ast}\mid/H_{1}^{\ast}$ on both sides of $H_{1}^{\ast}$.  The lines are fits with slopes $z\nu\approx 0.73$, as shown.}
  \label{fig:7per-scaling}
\end{figure}

\vspace*{12pt}
\noindent\textbf{\large{{Supplementary Information References}}}

\noindent [S1] Li, L., Checkelsky, J.G., Komiya, S., Ando, Y. \& Ong, N.P. Low-temperature vortex liquid in La$_{2-x}$Sr$_x$CuO$_4$.  \textit{Nature Phys.} \textbf{3}, 311--314 (2007).\\
\noindent [S2] Ando, Y. \textit{et al.} Resistive upper critical fields and irreversibility lines of optimally doped high-$T_c$ cuprates.  \textit{Phys. Rev. B} \textbf{60}, 12475--12479 (1999).\\
\noindent [S3] Sasagawa, T. \textit{et al.} Magnetization and resistivity measurements of the first-order vortex phase transition in (La$_{1-x}$Sr$_x)_2$CuO$_4$.  \textit{Phys. Rev. B} \textbf{61}, 1610--1617 (2000).\\
\noindent [S4] Le Doussal, P. Novel phases of vortices in superconductors.  \textit{Int. J. Mod. Phys. B} \textbf{24}, 3855--3914 (2010).\\
\noindent [S5] Houghton, A., Pelcovits, R.A. \& Sudb\o, A. Flux lattice melting in high-$T_c$ superconductors.  \textit{Phys. Rev. B.} \textbf{40}, 6763--6770 (1989).\\
\noindent [S6] Blatter, G., Feigel'man, M.V., Geshkenbein, V.B., Larkin, A.I. \& Vinokur, V.M. Vortices in high-temperature superconductors.  \textit{Rev. Mod. Phys.} \textbf{66}, 1125--1388 (1994).\\
\noindent [S7] Lemberger, T.R., Hetel, I., Tsukada, A., Naito, M. \& Randeria, M. Superconductor-to-metal quantum phase transition in overdoped La$_{2-x}$Sr$_x$CuO$_4$.  \textit{Phys. Rev. B} \textbf{83}, 140507(R) (2011).\\
\noindent [S8] Rosenstein, B. \& Li, D. Ginzburg-Landau theory of type II superconductors in magnetic field.  \textit{Rev. Mod. Phys.} \textbf{82}, 109--168 (2010).\\
\noindent [S9] Fisher, M.P.A., Grinstein, G. \& Girvin, S.M.  Presence of quantum diffusion in two dimensions: Universal resistance at the superconductor-insulator transition.  \textit{Phys. Rev. Lett.} \textbf{64}, 587--590 (1990).\\
\noindent [S10] Gantmakher, V.F. \& Dolgopolov, V.T. Superconductor-insulator quantum phase transition.  \textit{Physics-Uspekhi} \textbf{53}, 1--49 (2010).\\
\noindent [S11] Dobrosavljevi\'c, V., Trivedi, N. \& Valles, J.M.  Conductor-Insulator Quantum Phase Transitions (Oxford University Press, Oxford, 2012).\\
\noindent [S12] Leridon, B. \textit{et al.} Double criticality in the magnetic field-driven transition of a high-$T_c$ superconductor.  Preprint at http://arxiv.org/abs/1306.4583 (2013).\\
\noindent [S13] Bilbro, L. S. \textit{et al.} Temporal correlations of superconductivity above the transition temperature in La$_{2-x}$Sr$_x$CuO$_4$ probed by terahertz spectroscopy.  \textit{Nature Phys.} \textbf{7}, 298--302 (2011).\\
\noindent [S14] Li, L. \textit{et al.} Diamagnetism and Cooper pairing above $T_c$ in cuprates.  \textit{Phys. Rev. B} \textbf{81}, 054510 (2010).\\
\noindent [S15] Wang, Y. \textit{et al.} Onset of the vortexlike Nernst signal above $T_c$ in La$_{2-x}$Sr$_x$CuO$_4$ and Bi$_2$Sr$_{2-y}$La$_y$CuO$_6$.  \textit{Phys. Rev. B} \textbf{64}, 224519 (2001).\\
\noindent [S16] Rullier-Albenque, F., Alloul, H. \& Rikken, G. High-field studies of superconducting fluctuations in high-$T_c$ cuprates: Evidence for a small gap distinct from the large pseudogap.  \textit{Phys. Rev. B} \textbf{84}, 014522 (2011).\\
\noindent [S17] Shi, X. \textit{et al.} Emergence of superconductivity from the dynamically heterogeneous insulating state in La$_{2-x}$Sr$_x$CuO$_4$.  \textit{Nature Mater.} \textbf{12}, 47--51 (2013).

\end{document}